\documentclass[prb,onecolumn,amsmath,showkeys]{revtex4}

\usepackage{graphicx}
\usepackage{dcolumn}
\usepackage{bm}
\usepackage[T1]{fontenc}

\begin{document}

\title{Free energies of ferroelectric crystals from a microscopic approach}

\author{Gr\'egory Geneste$^1$}
\address{$^1$ Laboratoire Structures, Propri\'et\'es et Mod\'elisation des Solides, CNRS-UMR 8580,
Ecole Centrale Paris, Grande Voie des Vignes, 92295 Ch\^atenay-Malabry Cedex, France\\}

\begin{abstract}
The free energy of barium titanate is computed around the Curie temperature as a function of polarization $\vec P$ from the first-principles derived effective hamiltonian of Zhong, Vanderbilt and Rabe [Phys. Rev. Lett. {\bf 73}, 1861 (1994)], through Molecular Dynamics simulations coupled to the method of the Thermodynamic Integration. The algorithms used to fix the temperature (Nos\'e-Hoover) and/or the pressure/stress (Parrinello-Rahman), combined with fixed-polarization molecular dynamics, allow to compute a Helmholtz free energy (fixed volume/strain) or a Gibbs free energy (fixed pressure/stress). The main feature of this approach is to calculate the gradient of the free energy in the 3-D space ($P_x$,$P_y$,$P_z$) from the thermal averages of the forces acting on the local modes, that are obtained by Molecular Dynamics under the constraint of fixed $\vec P$. This work extends the method presented in [Phys. Rev. B {\bf 79} (2009), 064101] to the calculation of the Gibbs free energy and presents new features about the computation of the free energy of ferroelectric crystals from a microscopic approach. A careful analysis of the states of constrained polarization is performed at T=280 K ($\approx$ 15-17 K below T$_c$) especially at low order parameter. These states are found reasonably homogeneous for small supercell size (L=12 and L=16), until inhomogeneous states are observed at low order parameter for large supercells (L=20). The effect of this evolution towards multidomain configurations on the mean force and free energy curves is shown. However, for reasonable supercell sizes (L=12), the free energy curves obtained are in very good agreement with phenomenological Landau potentials of the litterature and the states of constrained polarization are homogeneous. Moreover, the free energy obtained is quite insensitive to the supercell size from L=12 to L=16 at T=280 K, suggesting that interfacial contributions, if any, are negligible at these sizes around T$_c$. The method allows a numerical estimation of the free energy barrier separating the paraelectric from the ferroelectric phase at T$_c$ ($\Delta G \approx$ 0.012-0.015 meV/5-atom cell). However, our tests evidence phase separation at low temperature and low order parameter, in agreement with the results of Tröster {\it et al} [Phys. Rev. B {\bf 72} (2005), 094103]. Finally, the natural decomposition of the forces into onsite, short-range, dipole-dipole and elastic-local mode interaction allows to make the same decomposition of the  free energy. Some parts of this decomposition can be directly calculated from the coefficients of the Effective Hamiltonian. 
\end{abstract}


\maketitle

\section{Introduction}
Two main theoretical approaches are currently used in the description of ferroelectricity in perovskite oxides. The first one, that we will call the {\it microscopic approach}, is based on first-principles calculations and effective hamiltonians. These effective hamiltonians are simplifications of the energy landscape of ferroelectric solids in terms of relevant degrees of freedom (local modes, displacement modes, homogeneous strain and, for more complex perovskites, antiferrodistortive modes\cite{afd}). They have even been recently extended to describe multiferroics by accounting for magnetic moments as degrees of freedom\cite{multiferro}.
Their mathematical form contains adjustable parameters that are usually derived from density-functional calculations. The effective hamiltonians, once constructed, can be solved either within Monte Carlo (MC) or Molecular Dynamics (MD) simulations. The physical quantities (polarization, strain...) are obtained as thermal averages over the equilibrium trajectories constructed from these two methods, that should provide equivalent sampling of phase space (MD) or configuration space (MC). These effective hamiltonians contain most of the thermodynamics of ferroelectric crystals and, in most cases, describe very well their evolution with temperature and pressure/stress with or without external electric field. They have been extended to treat low-dimensional systems (thin films, nanowires\cite{ponomareva}, dots\cite{fu2003,naumov2005}) under various mechanical and electrical boundary conditions. The first ab initio-derived effective hamiltonian has been constructed by Zhong, Vanderbilt and Rabe\cite{zhong94,zhong95} and successfully applied to barium titanate.

The second approach, that we will call the {\it macroscopic approach}, or phenomenological approach, is based on the Landau theory of phase transitions. This theory uses as central concept an incomplete thermodynamic potential (that can be a Helmholtz potential $F$ or a Gibbs potential $G$, or the thermodynamic potential of any other ensemble), commonly called Landau free energy, which is a function of the polarization $\vec P$ (order parameter) and can be defined from the restriction of the true thermodynamic potential to the microscopic states having a given polarization $\vec P$\cite{iniguez2001}.

In the canonic ensemble, it writes:
\begin{equation}
\nonumber
\tilde{F}(N,V,T;\vec P) = -k_B T ln \tilde{Z}(N,V,T;\vec P),
\end{equation}
with 
\begin{equation}
\nonumber
\tilde{Z}(N,V,T;\vec P) = \sum_{i/\vec P} e^{-\epsilon_i/k_B T},
\end{equation}
in which the incomplete canonical partition function defines a Helmholtz free energy $\tilde{F}$. The sum in this expression is over all the microstates $i$ having a polarization $\vec P$.
Within this definition, the incomplete free energy $\tilde{F}(\vec P)$ appears as directly related to the density of probability of the order parameter $\Pi(\vec P)$ by\cite{radescu}:
\begin{equation}
\label{probability}
\Pi(\vec P) = \Pi(\vec 0) e^{- [\tilde{F}(\vec P) - \tilde{F}(\vec 0)]/ k_B T}
\end{equation}

The same definitions in the isothermal-isobaric ensemble allow to define an incomplete Gibbs free energy:
\begin{equation}
\nonumber
\tilde{G}(N,P,T;\vec P) = -k_B T ln \tilde{\Delta}(N,P,T;\vec P),
\end{equation}
with 
\begin{equation}
\nonumber
\tilde{\Delta}(N,P,T;\vec P) = \sum_{i/\vec P} e^{-(\epsilon_i + PV_i)/k_B T},
\end{equation}

the (incomplete) isothermal-isobaric partition function.

The Landau approach describes quite well the thermodynamics of many ferroelectrics. It was applied in a pioneering work by Devonshire on bulk barium titanate a long time ago\cite{devonshire49,devonshire51,devonshire54} and has been applied since then on many other ferroelectric materials and in many other forms than bulk\cite{landau_book}. In Landau theory, the thermodynamic potential is expanded in power series of the polarization $\vec P$ (the order parameter in ferroelectric crystals), usually up to 6$^{th}$ or 8$^{th}$ order. The quadratic coefficient is assumed to vary linearly with temperature while the higher-order coefficients are usually assumed as constants (more complex potentials with temperature-dependent high-order coefficients have been recently proposed\cite{bell_cross,wang2007}). These free energies are used to model the thermodynamics of the system considered around the critical temperature and even further, even though the Landau approach is supposed to break down near the critical temperature. Since such polynomial developments are by nature approximate, it appears legitim to try to use the microscopic models introduced hereabove to get insight into the Landau potentials by using standard numerical simulation tools.


However, the connection between the two approaches is not simple. Indeed, on the one hand, the MC or MD methods are unable to give directly access to the thermodynamic potential, that can not be computed as the thermal average of some microscopic quantity. Indirect methods can nevertheless be employed: using the fact that the Landau free energy is directly related to the density of probability of the order parameter, it may be possible to compute the free energy from the probability distribution obtained from a long enough MC or MD run\cite{iniguez,radescu}. However, if such calculations are possible and efficient in the simple case of the $\phi^4$ model\cite{radescu}, it is not the case of ferroelectric crystals described by the effective hamiltonians introduced above, for which the high-free energy parts (for instance the part of $\tilde{F}$ at low order parameter in the ferroelectric phase) are not sampled correctly (or even not at all), making the free energy accessible only in the vicinity of the equilibrium value of the order parameter\cite{iniguez}.

A clever alternative method was proposed by I$\tilde{n}$iguez {\it et al}\cite{iniguez2001} to calculate from MC simulations the coefficients of the Landau free energy expansion, that consisted in applying an external electric field in order to displace the minimum of the free energy in the ($P_x,P_y,P_z$) space. Recently, we proposed to use the method of the thermodynamic integration\cite{geneste1}, coupled to MD simulations, to access numerically to the free energy as a function of polarization (calculated as a Helmholtz free energy, under constant strain tensor). This method allows a correct sampling of the high-free energy regions (where the "reaction coordinate" has a low probability) since it consists of MD runs under the constraint of fixed reaction coordinate. Thermodynamic integration is a technique commonly applied in computational chemistry to obtain free energy profiles along a reaction coordinate. In the present case, the order parameter (polarization) is taken as the reaction coordinate. It has also already been applied in the framework of Landau theory to study the cubic-tetragonal phase transition in ZrO$_2$ by Fabris {\it et al}\cite{fabris2001}. Other techniques such as the Wang-Landau method can also be used\cite{troster2005}.

On the other hand, the computation of Landau potentials is delicate for various theoretical reasons. Indeed, fixing the order parameter to a given value ({\it i.e.} sampling the phase space under the constraint of fixed order parameter) can raise unexpected problems, especially at low value, where the system might evolve to inhomogeneous states and separate into "domains" below T$_c$. In the case where this phase separation occurs, changing the order parameter from zero to higher values simply results, at least at low order parameter, in some domain wall motion (or more complex interface motions and reorganizations), and the computed free energy does probably not correspond to the free energy expected as the central concept of Landau theory, in which sufficiently homogeneous states are assumed all along the free energy curve. Tröster {\it et al}\cite{troster2005} have studied this phenomenon in details in the framewotk of the so-called $\phi^4$ model through Wang-Landau computations of the free energy. Moreover, if surface or interfacial contributions enter the computation of the thermodynamic potential, the calculated object might not be a volume quantity and not have the extensivity required by such a thermodynamic function.

Essentially, such problems are related to the basic fact that defining properly the polarization as a continuous physical quantity requires the choice of a spatial averaging length\cite{troster2005} (as it is the case, in fact, for all macroscopic quantities used in electrodynamics). This averaging length scale $L$ should be chosen below the correlation length $\xi$ so that the local order parameters (electric dipoles in our case) remain correlated within the averaging volume and that each dipole only fluctuates around the averaged value all over the averaging volume. Conversely, it should be also chosen much larger than the lattice parameter $a$ (elementary distance between first neighbor local order parameters) so that the order parameter as a slowly varying macroscopic quantity keeps a sense. For these reasons, it is admitted that the Landau free energy should be defined with respect to a given averaging length $L$ (coarse-graining length)\cite{troster2005}, that should fullfil the condition $a << L << \xi$.

The correlation length is expected to decrease with temperature above T$_c$ and to increase with it below T$_c$. However, the notion of correlation volume/length is quite complex in ferroelectric systems\cite{lines72,corrvol}, due to the peculiar nature of the dipole-dipole interaction, that is {\it long-ranged} and leads to strongly anisotropic correlations. As a consequence, it is the shape rather than the volume that defines a region within which dipolar correlations can produce ferroelectric order\cite{corrvol}. Lines\cite{lines72} pointed out more than thirty years ago the difficulty to define properly a correlation length in a ferroelectric. Anyway, we will assume in this paper that the notion of correlation length/volume keeps a sense in ferroelectric solids, that will help us to understand from a qualitative point of view the free energies computed with various supercell sizes and for various temperatures.

From a numerical point of view, we use as averaging volume the supercell used for simulating the bulk system within periodic boundary conditions ($L^3$). Computing a relevant free energy as a function of order parameter would therefore require the supercell size to be lower than $\xi$. If such a condition is not fullfilled, inhomogeneous configurations are expexted to appear at low order parameter and low temperature.
Since $\xi$ decreases rapidly when the temperature decreases below T$_c$, and that, of course, a too small supercell can not be used to prevent from too large finite-size effects, it suggests that computing a relevant free energy within the present method is possible only above a certain temperature in the ferroelectric phase. However, the identification of such a free energy with the Landau free energy is still a controversial issue\cite{troster2005}. In the same manner, at a given temperature, phase separation is expected to occur at low order parameter for large enough supercells of size $L > \xi$. We add that even when $L < \xi$ (around the critical temperature for instance, where $\xi$ can reach high values), the computation suffers from finite-size effects, unavoidable in that case within periodic boundary conditions.

In the present work, we present some features that may help to get insight into Landau potentials of ferroelectric materials from a microscopic approach, at least around the Curie temperature. We extend the approach of Ref.~\onlinecite{geneste1} to MD simulations under constant pressure, that allow to compute a Gibbs free energy as a function of polarization. Once applied to BaTiO$_3$, the free energy curves obtained are very similar to what is expected from Landau theory of first-order phase transitions and, after fitting these curves by 8$^{th}$ order polynomial functions, an excellent agreement is found with availables phenomenological Landau potentials of the litterature.

We also conduct a very careful analysis of the states of constrained order parameter for the lowest temperature studied ($\approx$ T$_c$ - 15 K). They are found to be more and more inhomogeneous as the supercell size increases. For the largest supercell (L=20), phase separation starts to occur at this temperature. Below this size (between L=12 and L=20), the mean force and free energy curves obtained are independent on the supercell size, showing that the computed free energy is free of interfacial contribution and is actually a volume quantity. For a reasonable supercell size (L=12), the macroscopic states of constrained polarization are shown to be homogeneous (this is carefully tested by increasing the number of time steps). However, we show that below a given temperature ($\approx$ 240-250 K) the use of a 12 $\times$ 12 $\times$ 12 supercell leads to inhomogeneous states at low order parameter, as expected.


We also discuss the contributions to the free energy of a ferroelectric crystal in terms of the four contributions (onsite, short-range, dipole-dipole and elastic-local mode interaction) that constitute the effective hamiltonian. We show that two of them are independent on the temperature when expressed as functions of the mean local mode $\vec u$ (dipole) instead of polarization $\vec P$.

\section{Theoretical and computational details}

\subsection{Hamiltonian}
We have used the Effective Hamiltonian of Zhong, Vanderbilt and Rabe\cite{zhong94,zhong95}, devoted to BaTiO$_3$. This hamiltonian uses as degrees of freedom the "local modes" $\vec u_i$ (local order parameters), that roughly represent the local dipoles that do exist instantaneously in each unit cell, the (mechanical) displacement modes $\vec v_i$, from which the inhomogeneous strain tensor $\left\{\eta^I_l(i)\right\}$ is constructed for each unit cell $i$, and the homogeneous strain tensor $\left\{\eta^H_l\right\}$. The strain tensor at cell $i$ is thus given by $\eta_l(i)=\eta^H_l+\eta^I_l(i)$.

This hamiltonian $H^{eff}(\left\{\vec u_i\right\}, \left\{\eta_l(i)\right\})$ consists of several terms, among which a {\it long-range} interaction is included to model the dipole-dipole interaction (this term is absolutely necessary to provide a realistic simulation of a ferroelectric system):

\begin{equation}
H^{eff}(\left\{\vec u_i\right\}, \left\{\eta_l(i)\right\}) = E^{self}(\left\{\vec u_i\right\}) + E^{dpl}(\left\{\vec u_i\right\}) + E^{short}(\left\{\vec u_i\right\}) + E^{elas}(\left\{\eta_l(i)\right\}) + E^{int}(\left\{\vec u_i\right\}, \left\{\eta_l(i)\right\})
\end{equation}

The first term $E^{self} = \sum_i E(\vec u_i)$ is local and consists of a fourth-order polynomial function in the $\vec u_i$:

\begin{equation}
E(\vec u_i) = \kappa_2 \left\|\vec u_i\right\|^2 + \alpha \left\|\vec u_i\right\|^4 + \gamma(u_{ix}^2 u_{iy}^2 + u_{iy}^2 u_{iz}^2 + u_{ix}^2 u_{iz}^2)
\end{equation}

The second term $E^{dpl}$ is crucial to the description of ferroelectric systems: it is the (long-range) dipole-dipole interaction between the local modes:
\begin{equation}
E^{dpl} = \frac{{Z^{*}}^2}{\epsilon_{\infty}} \sum_{i<j} \frac{\vec u_i . \vec u_j - 3(\hat{\vec R}_{ij}.\vec u_i)(\hat{\vec R}_{ij}.\vec u_j)}{R_{ij}^3},
\end{equation}
with $\hat{\vec R}_{ij} = \vec R_{ij} / R_{ij}$, $\vec R_{ij}$ being the lattice vector joining cell $i$ to cell $j$. $Z^{*}$ is the effective charge associated to the local modes and $\epsilon_{\infty}$ the electronic dielectric constant. This long-range interaction can be expressed as:
\begin{equation}
E^{dpl} = \sum_{i,j,\alpha,\beta} Q_{ij,\alpha \beta} u_{i \alpha} u_{j \beta},
\end{equation}
the $Q$ matrix being calculated by using the well-known Ewald summation method.

The third term describes short-range interactions between neighboring local modes (up to the third neighbor):
\begin{equation}
E^{short} = \frac{1}{2} \sum_{i\neq j} \sum_{\alpha,\beta} J_{ij, \alpha \beta} u_{i \alpha} u_{j \beta}
\end{equation}

The fourth term is the elastic energy associated to the strain. This term is calculated from the three elastic constants B$_{11}$, B$_{12}$ and B$_{44}$ of the parent cubic phase.

Finally the fifth term describes the coupling between the local modes and the strain. In ferroelectric systems, this coupling is at the origin of the so-called electrostrictive effects and is of primary importance. It is supposed to be local and takes the following form:
\begin{equation}
E^{int} = \frac{1}{2} \sum_{i} \sum_{l \alpha,\beta} B_{l \alpha \beta} \eta_l(i) u_{i \alpha} u_{j \beta}
\end{equation}

All the parameters of this hamiltonian are determined from first-principles calculations of well-chosen configurations\cite{zhong94,zhong95} in the framework of the Local Density Approximation (LDA) to Density Functional Theory (DFT). The reader is refered to Ref.~\onlinecite{zhong95} for their precise value.
This effective hamiltonian describes very well the thermodynamics of barium titanate, especially its complex sequence of phase transitions (rhombohedral - orthorhombic - tetragonal - cubic), with a Curie temperature however a little too low (around 300 K).

\subsection{Numerical method}
We perform MD simulations based on this hamiltonian\cite{md,geneste1}. According to the algorithm used, we can perform either fixed temperature simulations (Nos\'e-Hoover algorithm\cite{nose84,hoover85}), fixed stress tensor simulations (Parrinello-Rahman algorithm\cite{pr}) or both. The combination of the two allows to reproduce with a very good accuracy the sequence of phase transitions and temperature evolution of strain and polarization of barium titanate\cite{md}, as found from MC simulations\cite{zhong94,zhong95}. The equations of motion in the Parrinello-Rahman scheme are recalled hereafter:

\begin{equation}
\nonumber
m_{lm}\frac{d^2 \vec u'_i}{dt^2} =  H(t)^{-1}.\vec f_i^{lm} - m_{lm} G(t)^{-1}.\frac{dG}{dt}(t).\frac{d \vec u'_i}{dt}
\end{equation}

\begin{equation}
\nonumber
m_{dsp}\frac{d^2 \vec v'_i}{dt^2} =  H(t)^{-1}.\vec f_i^{dsp} - m_{dsp} G(t)^{-1}.\frac{dG}{dt}(t).\frac{d \vec v'_i}{dt},
\end{equation}
respectively for the local modes ($\vec u_i = H(t).\vec u'_i$) and diplacement modes ($\vec v_i = H(t).\vec v'_i$). $m_{lm}$ and $m_{dsp}$ are respectively the masses associated to the local modes and to the displacements modes. $H(t)$ (the $3 \times 3$ matrix formed by the components of the three vectors $\vec a$, $\vec b$ and $\vec c$ that define the supercell) evolves according to:

\begin{equation}
\nonumber
W \frac{d^2 H}{dt^2}(t) = (\underline{\sigma_{0}} - \underline{\sigma}).\underline{\omega},
\end{equation}

in which $W$ is a "mass" associated to the dynamics of the supercell vectors $\vec a$, $\vec b$ and $\vec c$, which has to be correctly chosen. $\underline{\omega}_{ij} = \partial \Omega / \partial H_{ij}$ ($\Omega$ is the volume of the supercell) and $G(t) = ^tH(t).H(t)$. $\underline{\sigma}$ and $\underline{\sigma_{0}}$ are respectively the instantaneous stress tensor and the desired stress tensor.

In the preliminary work of Ref.~\onlinecite{geneste1}, we have performed MD simulations under {\it fixed polarization}\cite{geneste1}. These simulations are achieved by adding external forces in the time-evolution equations of motion of the local modes (see Ref.~\onlinecite{geneste1} for a precise explanation):

\begin{equation}
\nonumber
m_{lm}\frac{d^2 u_{i,\alpha}}{dt^2} = f_{i,\alpha}^{lm} - \zeta_{\alpha},
\end{equation}

with $\zeta_{\alpha} = 1/N \sum_i f_{i,\alpha}^{lm}$. $u_{i,\alpha}$ is the $\alpha$-component of the $i^{th}$ local mode (which has the dimension of a displacement), $m_{lm}$ is the mass associated to the local modes, $f_{i,\alpha}^{lm}$ the $\alpha$-component of the force acting on the $i^{th}$ local mode and $N$ the number of 5-atom unit cells in the supercell used for the simulation. The quantity $\vec \zeta = (\zeta_x, \zeta_y, \zeta_z)$, calculated in Ref.~\onlinecite{geneste1} as a Lagrange multiplier, is formally equivalent to an external electric field that would be applied to the system and would maintain invariant the polarization. The total instantaneous external force applied to the system to maintain the polarization fixed is thus $- \sum_i \vec f_{i}^{lm}$. This Molecular Dynamics under constraint is assumed to provide a correct sampling of the subspace of phase space corresponding to a fixed value of the order parameter, defined by an average over the whole simulation box.

Based on such simulations, we have applied the method of the {\it thermodynamic integration}  to compute the difference of free energy $\Delta \tilde{F}(N,\left\{ \eta \right\},T;\vec u)= \tilde{F}(N,\left\{ \eta \right\},T;\vec u) - \tilde{F}(N,\left\{ \eta \right\},T;\vec u=\vec 0)$ between $\vec u= \vec 0$ and $\vec u$. We use for convenience in the following the average local mode $\vec u = 1/N \sum_i \vec u_i$ as order parameter instead of the polarization $\vec P$ ($\vec P = N Z^{*}/\Omega \vec u$, $Z^{*}$ being the effective charge associated with the local modes and $\Omega$ is the volume of the supercell).

This difference is obtained as the integral over a continuous path in the 3-D space ($u_x$,$u_y$,$u_z$) of minus the thermal average of the total forces acting on the local modes, this thermal average being computed under fixed volume (more precisely under fixed strain tensor), fixed temperature and fixed mean local mode:

\begin{equation}
\label{thermo_int}
\tilde{F}(N,\left\{ \eta \right\},T;\vec u) - \tilde{F}_0(N,\left\{ \eta \right\},T) = - \oint \sum_{i} \left\langle \vec f^{lm}_i \right\rangle_{N,\left\{ \eta \right\},T;\vec u'} . d\vec u',
\end{equation}

The subscript $0$ in this formula and in the following is used to denote the quantities at $\vec u = \vec 0$, and the integral is performed over a continuous path joining $\vec u = \vec 0$ to $\vec u$ in the 3-D space $(u_x,u_y,u_z)$. A complete and rigorous proof of this equation can be found in Ref.~\onlinecite{geneste1} in terms of partition functions, but basically it can be recovered from the principles of thermodynamics: indeed under fixed volume $V$ and fixed temperature $T$, the infinitesimal variation of the free energy $d\tilde{F}$ when the polar displacements evolve from $\vec u$ to $\vec u + d\vec u$ is equal to $\delta W'$, the work of the external forces (other than pressure forces), {\it i.e.} in our case the work of the external force $- N \left\langle \vec \zeta \right\rangle = - \sum_i \left\langle \vec f_{i}^{lm} \right\rangle$:

\begin{equation}
\label{helmholtz1}
d\tilde{F}(\vec u) = \delta W' = - \sum_i \left\langle \vec f_{i}^{lm} \right\rangle_{N,\left\{ \eta \right\},T;\vec u}. d\vec u
\end{equation}

which writes in a local form:

\begin{equation}
\label{mean_force}
\sum_{i} \left\langle \vec f^{lm}_i \right\rangle_{N,\left\{ \eta \right\},T;\vec u} = - \vec \nabla_{\vec u} \tilde{F}(N,\left\{ \eta \right\},T;\vec u)  
\end{equation}

and provides by integration Eq.~\ref{thermo_int}. The free energy is thus computed as a potential of mean force (PMF). Its differential $d \tilde{F}$ when the system evolves reversibly from $\vec u $ to $\vec u + d \vec u$ is the infinitesimal work that has to be provided to the system (or received from it) to allow such a {\it reversible} transformation. This work is that of the external forces added in the equations of motion of the local modes\cite{md}.

The extension to the Gibbs free energy is straightforward\footnote{the thermodynamic integration providing the Gibbs free energy is formally identical to that providing the Helmholtz free energy. This is due to the peculiar form of the Effective Hamiltonian, in which the local modes and the homogeneous strain are independent variables.}:

\begin{equation}
\label{gibbs1}
\Delta \tilde{G}(N,\left\{ \sigma \right\},T;\vec u) = \tilde{G}(N,\left\{ \sigma \right\},T;\vec u) - \tilde{G}_0(N,P,T)  = - \oint \sum_{i} \left\langle \vec f^{lm}_i \right\rangle_{N,\left\{ \sigma \right\},T;\vec u'} . d\vec u',
\end{equation}

in which the thermal averages are computed under fixed temperature, stress tensor and polarization. This yields:

\begin{equation}
\tilde{G}(N,\left\{ \sigma \right\},T;\vec P) - \tilde{G}_0(N,\left\{ \sigma \right\},T) = - \oint \frac{\Omega (\vec P')}{NZ^{*}} \sum_{i} \left\langle \vec f^{lm}_i \right\rangle_{N,\left\{ \sigma \right\},T;\vec P'} . d\vec P'
\end{equation}

In this last expression, the relaxation of strain inducing a volume variation ($\Omega(P,T,\vec u)$) along the path is taken into account.

\subsection{Computational details}
In this work, we have computed $\Delta \tilde{G}(N,\left\{ \sigma \right\},T;\vec u) = \tilde{G}(N,\left\{ \sigma \right\},T;\vec u) - \tilde{G}_0(N,\left\{ \sigma \right\},T)$ for BaTiO$_3$, for 6 temperatures around the Curie temperature along the [100] direction. This is achieved through the following procedure. The calculation is performed every $\Delta u =$ 0.001 a$_0$, starting from $\vec u = \vec 0$, which is found to be a fine enough grid to make the integration. Except in Sec.~\ref{tests} in which the effect of the supercell size is systematically tested, we use a 12 $\times$ 12 $\times$ 12 supercell with periodic boundary conditions.

First we perform for each point 10$^5$ steps of MD by combining fixed pressure (Parrinello-Rahman algorithm), fixed temperature (Nos\'e-Hoover algorithm) and fixed polarization\cite{geneste1}. The (hydrostatic) pressure is fixed to -4.8 GPa, which is the value used by Zhong {\it et al} to correct the underestimation of the lattice constant of BTO within the Local Density Approximation\cite{zhong95}. The 50000 first steps are used to equilibrate the system. The last 50 000 are used to average the strain. The evolution of $\vec P$ computed at constant pressure and for a given temperature is indeed accompanied by an evolution of the strain, as can be seen on Fig.~\ref{figure0}.

Then, for each value of $\vec u$ of the grid, the calculation is restarted (for 10$^5$ steps) under fixed strain tensor (the one averaged over the previous simulation), polarization and temperature. The final 50 000 steps of this second run are used to calculate the thermal average of the forces acting on the local modes. We check at the end of this second run that the stress tensor (obtained from the average over the 50 000 last steps) corresponds to the hydrostatic pressure of -4.8 GPa (it works with an accuracy of $\approx$ 0.01 GPa). This procedure in two times is applied to ensure that the combination of the three algorithms (Nos\'e-Hoover, Parrinello-Rahman, fixed polarization) does not induce artefacts that would bias the statistical averages. In fact, the free energy profiles obtained from the forces averaged over the 50 000 last steps of the first run are almost identical to those obtained from the second one. We point out that at least 50 000 steps are necessary to compute properly the mean force (forces are microscopic quantities having very large fluctuations). The mean force in numerically integrated as in Refs.~\onlinecite{geneste1,geneste2}.

\section{States of constrained polarization at T=280 K ($\approx$ T$_c$ - 15-17 K)}
\label{tests}
In the method employed here, the free energy is computed from MD simulations under fixed polarization, as explained above. The result is a sampling of phase space under the constraint of fixed order parameter, defined from an average over the whole supercell. However, at low order parameter, the system might separate into domains or at least become inhomogeneous\cite{troster2005} inside the supercell.

Moreover, the existence of domains would yield a difficulty to define properly the free energy as a volume quantity, by incorporating surface or interfacial contributions in the computed free energy. In Sec.~\ref{homogeneity} we check the reasonable homogeneity of the states of constrained order parameter obtained in 12 $\times$ 12 $\times$ 12 supercells through 50000 time steps for the lowest temperature studied (280 K). In Sec.~\ref{simulation_time}, we test that this homogeneity is systematically improved by increasing the number of time steps of the simulation. The existence of interfacial contributions can be systematically tested by increasing the supercell size (Sec.~\ref{supercell}) or decreasing the temperature (Sec.~\ref{temperature}).

\subsection{Homogeneity of the macroscopic states of constrained order parameter}
\label{homogeneity}
\begin{figure}[htbp]
    {\par\centering
    {\scalebox{0.65}{\includegraphics{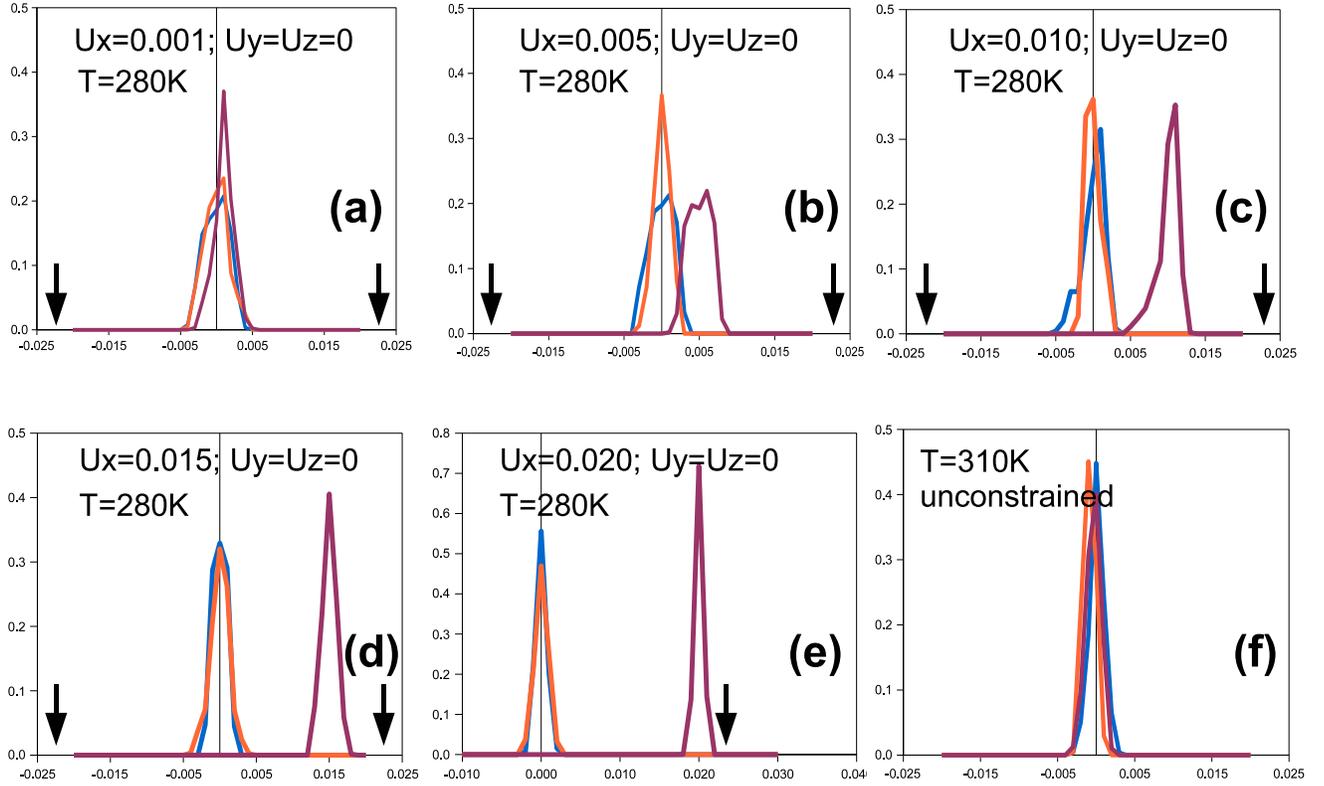}}}
    \par}
     \caption{{\small Distribution of the time-averaged components of the local modes (arb. units) (averaged over 50 000 steps), for 5 constrained values of the order parameter, at T=280 K (panels a,b,c,d and e). The purple (resp. blue and orange) lines refer to the $x$ (resp. $y$ and $z$) component. The graphs show that the macroscopic states of constrained polarization remain homogeneous along the simulation (no phase separation). The broadness of the peaks reflect fluctuations and can be systematically reduced (see Fig.~\ref{locmode2}). It can be compared to panel f, where the distribution of the local modes in an unconstrained system at T=310K ({\it i.e.} above T$_c$) is plotted after the same simulation time. The black arrows localize approximately the values of the spontaneous polarization at T=280 K (where the peaks are expected if phase separation occurs). The supercell is 12 $\times$ 12 $\times$ 12 and the local modes are in a$_0$ (lattice constants) units.}}
    \label{locmode}
\end{figure}

We first perform a series of simulation tests in a 12 $\times$ 12 $\times$ 12 supercell: we check that all along the simulated path, the system does not separate into domains, even for small values of the order parameter. We compute the time-averaged local mode {\it on each site of the simulation supercell}:

\begin{equation}
\vec u_i = \frac{1}{N_{st}} \sum_{k=1}^{N_{st}} \vec u_i(k),
\end{equation}

in which $\vec u_i (k)$ is the local mode at cell $i$ at time step $k$ ($N_{st}$ is the number of time steps), and plot their distribution (Fig.~\ref{locmode}). Clearly, a multidomain state would result in two distinct peaks at least for one component ($x$, $y$ or $z$). In all cases, we obtain single-peak distributions. On the graphs of Fig.~\ref{locmode}, we have put arrows to indicate approximately the values of the spontaneous polarization of tetragonal BTO at the temperature considered here (T=280 K). A two-domain state would result in two peaks approximately localized close to those arrows. In all the cases, the width of the local mode distribution is well below this spontaneous polarization.

\begin{figure}[htbp]
    {\par\centering
    {\scalebox{0.65}{\includegraphics{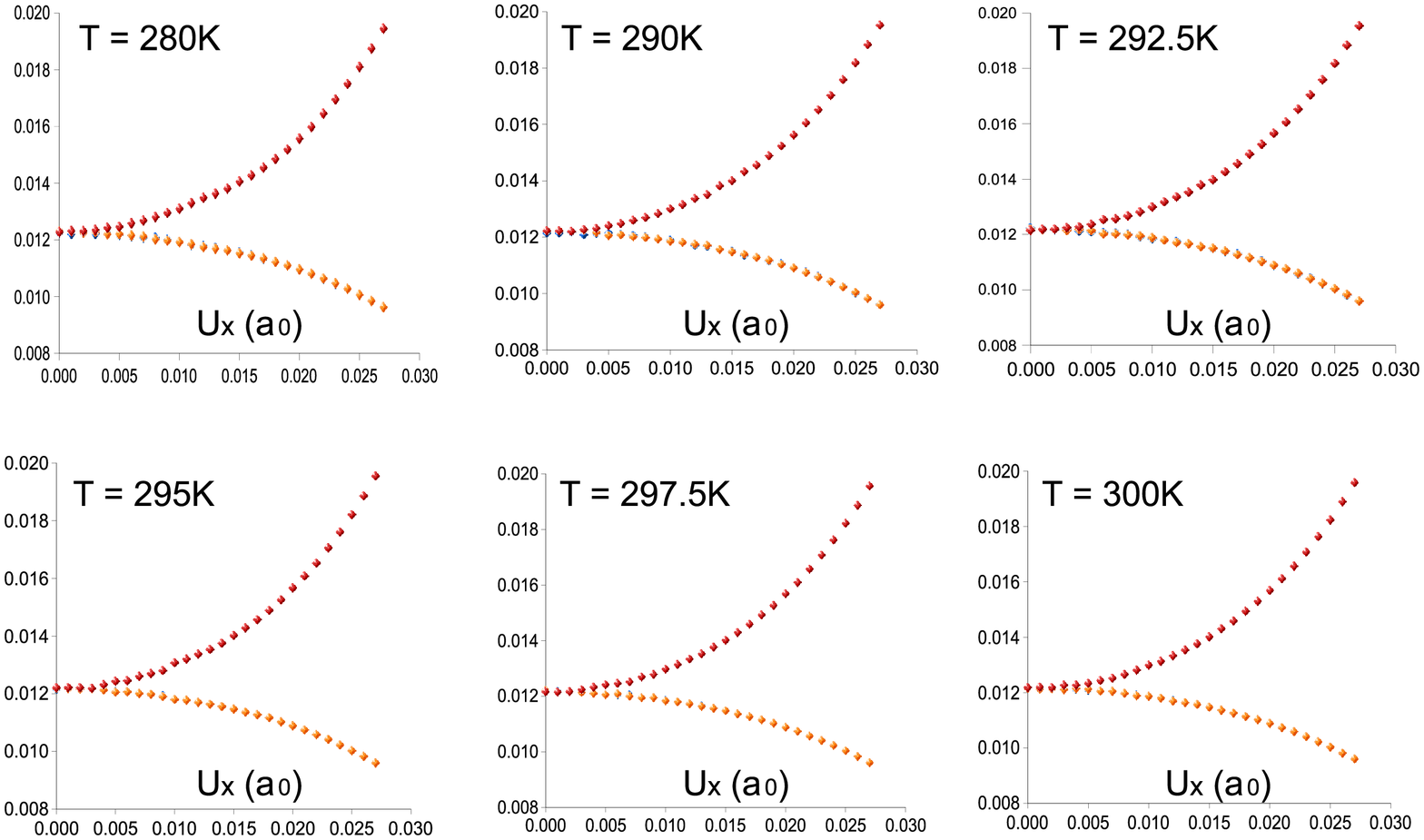}}}
    \par}
     \caption{{\small Diagonal components of homogeneous strain tensor $\eta_1$ (red diamonds), $\eta_2$ (orange diamonds) and $\eta_3$ (blue diamonds) as a function of the order parameter, for six temperatures around T$_c$. P = -4.8 GPa, the Parrinello-Rahman algorithm is used to fix the stress along the path.}}
    \label{figure0}
\end{figure}

The absence of ferroelectric domains is also obvious from the evolution of the strain tensor components along the simulated path, plotted on Fig.~\ref{figure0} for six temperatures around the Curie temperature: the evolution is smooth and continuous, with $\eta_1 = \eta_2 = \eta_3 \approx$ 0.012 for zero order parameter ($u_x = u_y = u_z = 0$). If ferroelectric domains were present, one component would jump to a high value corresponding roughly to the tetragonal strain at the corresponding temperature. Note that this 0.012 strain obtained at low order parameter is precisely the value that would be extrapolated from the paraelectric phase across T$_c$, suggesting that our constrained simulation for zero order parameter allows to reach some constrained paraelectric phase below the Curie temperature, which provides a physical picture fully compatible with Landau theory.

Thus, the computed free energy does probably not contain significant interfacial contribution (we show hereafter in Sec.~\ref{supercell} that the present free energy does not significantly vary when increasing the supercell size to 16 $\times$ 16 $\times$ 16). Anyway, this has to be controlled severely in each simulation. A quick guess on what is going on can be obtained by examining the strain tensor components evolution (similarly, if one computes a Helmholtz energy under fixed strain, by examining the stress).

Moreover, as clearly explained in Ref.~\onlinecite{troster2005}, if a multidomain state occurs at fixed low order parameter, the increase of the order parameter simply results in one domain growing at the expend of the other, or in a succession of more or less complex interface motions. As a consequence, as long as the domain walls move without interacting with each other, the free energy remains "flat" and yields very characteristic curves\cite{troster2005}. Such phase separation might occur at low temperature and low displaciveness\cite{troster2005}. Precisely, this does not occur in the present simulations (Fig.~\ref{figure1}), that are performed around the Curie temperature.

Anyway such cases occur at "low" temperature, {\it i.e.} a few 10 K below the phase transition (probably when the correlation length falls below the supercell size - see hereafter), or when the geometry enforces ferroelectricity (for instance we find it in thin films or with strongly distorted cells, for which a strong electrostrictive coupling favours a ferroelectric state).

\subsection{Simulation time}
\label{simulation_time}
\begin{figure}[htbp]
    {\par\centering
    {\scalebox{0.65}{\includegraphics{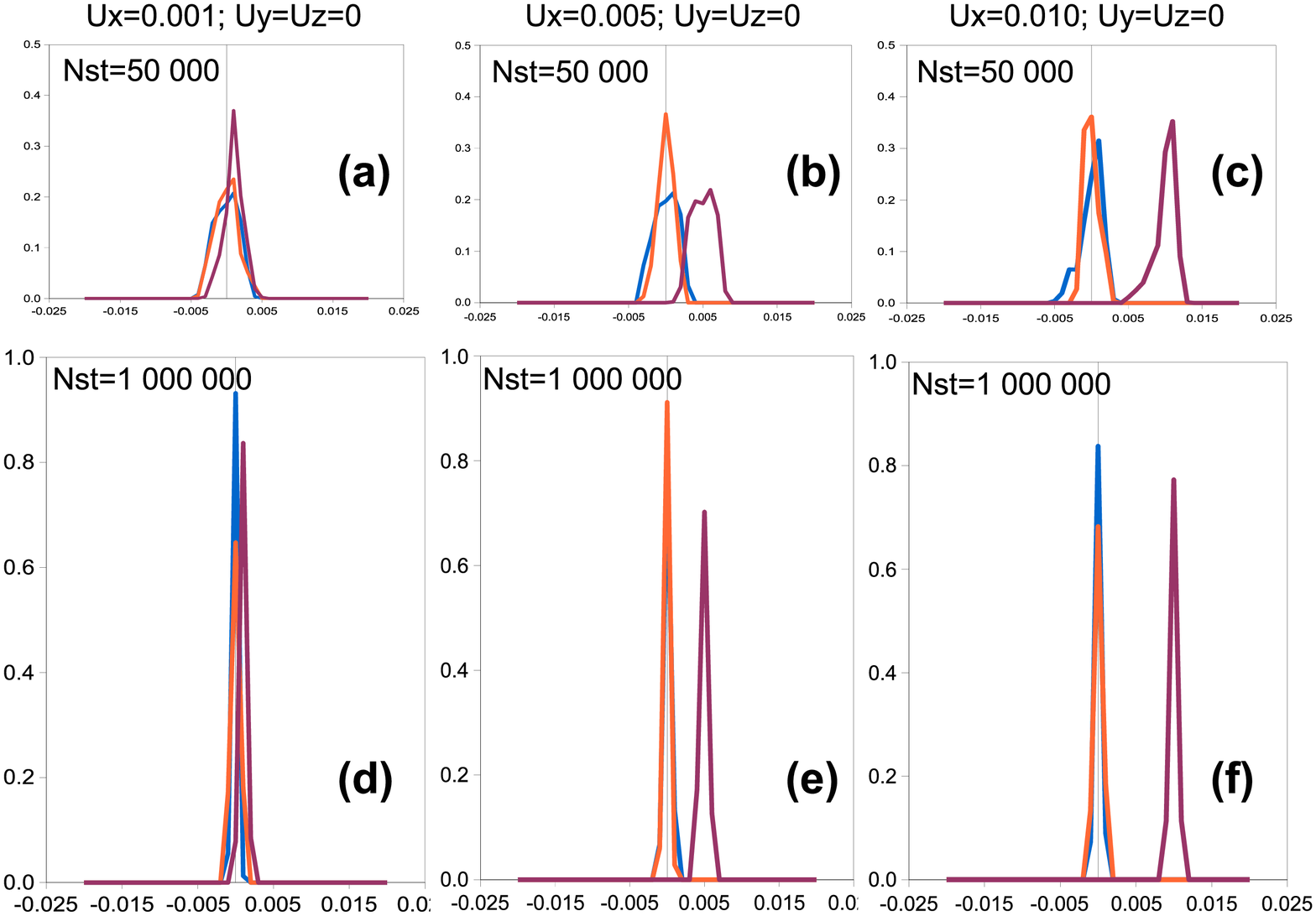}}}
    \par}
     \caption{{\small Distribution of the time-averaged components of the local modes (arb. units), for 3 constrained values of the order parameter, at T=280 K. Panels a,b,c: average over 50000 steps. Panels d,e,f: average over 1000000 steps. The purple (resp. blue and orange) lines refer to the $x$ (resp. $y$ and $z$) component. The supercell is 12 $\times$ 12 $\times$ 12 the local modes are in a$_0$ (lattice constants) units.}}
    \label{locmode2}
\end{figure}

\begin{figure}[htbp]
    {\par\centering
    {\scalebox{0.65}{\includegraphics{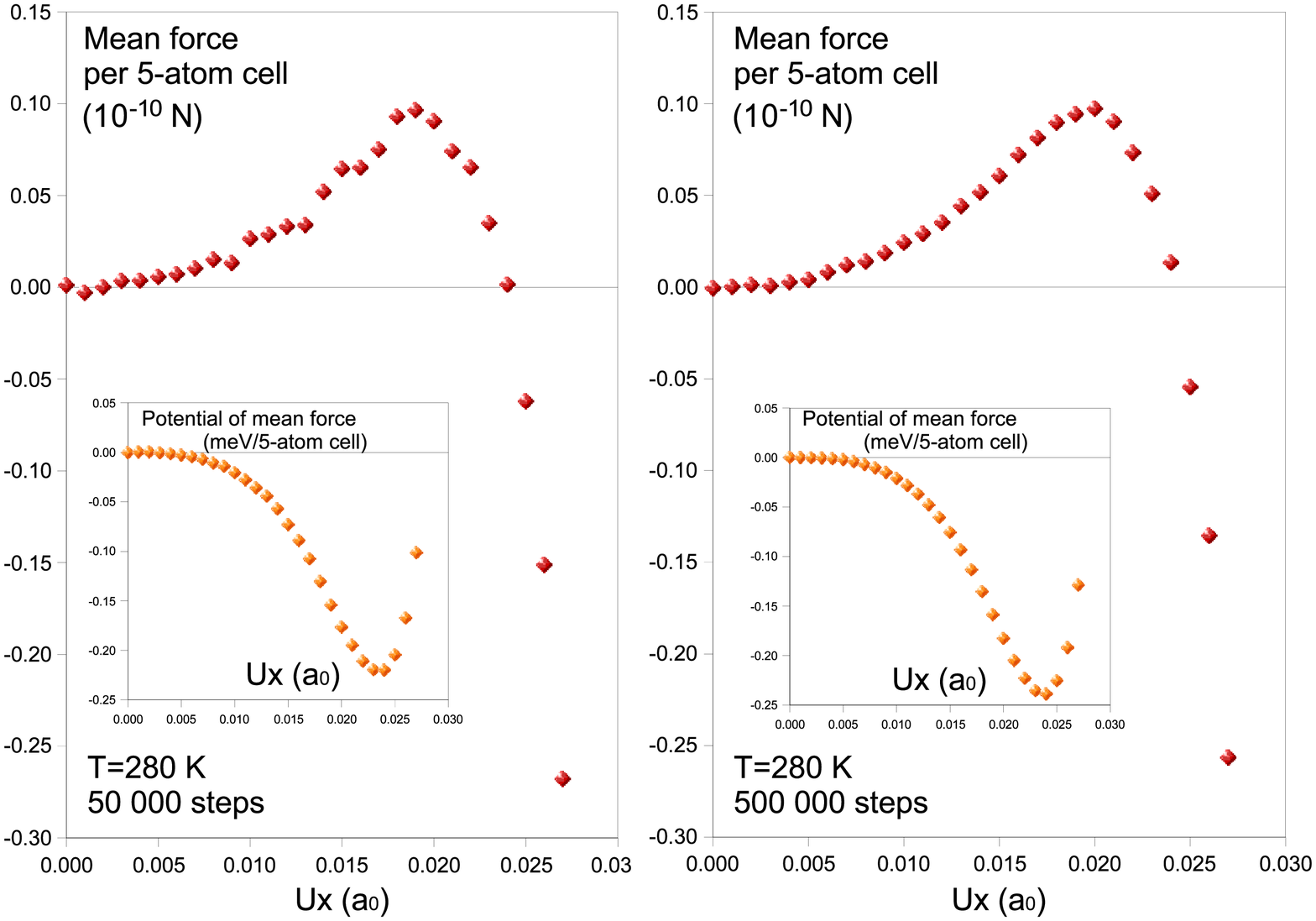}}}
    \par}
     \caption{{\small Mean force as a function of mean local mode (arb. units) and potential of mean force (obtained by integration of minus the mean force). (a): averaging over 50 000 steps; (b): averaging over 500 000 steps. Note that the two free energy curves have not exactly the same depth due to linear deviations as the integration is performed.}}
    \label{meanforce}
\end{figure}

The simulations are systematically improved by increasing the simulation time: Fig.~\ref{locmode2} shows that the time-averaged local mode distribution becomes more and more peaked as the simulation time is increased, showing that we are not sampling some metastable state. To emphasize this point, we plot on Fig.~\ref{meanforce} the mean force per 5-atom cell $\frac{1}{N} \sum_i \left\langle f_{ix}^{lm} \right\rangle$ and the free energy curve (obtained by integrating the mean force) computed at T=280 K after averaging over 50 000 steps and 500 000 steps. A systematic improvement is found. Any phase separation would result in plateaus in the mean force curve, which are not observed. Moreover, the average over 500 000 steps provides a very smooth and regular curve. We observe also that the potentials of mean force obtained in the two cases are almost identical, because the numerical errors on the forces tend to cancel when the integration is performed. Thus very good quality curves are obtained with an average over 50 000 steps.

\subsection{Effet of the supercell size}
\label{supercell}
\begin{figure}[htbp]
    {\par\centering
    {\scalebox{0.65}{\includegraphics{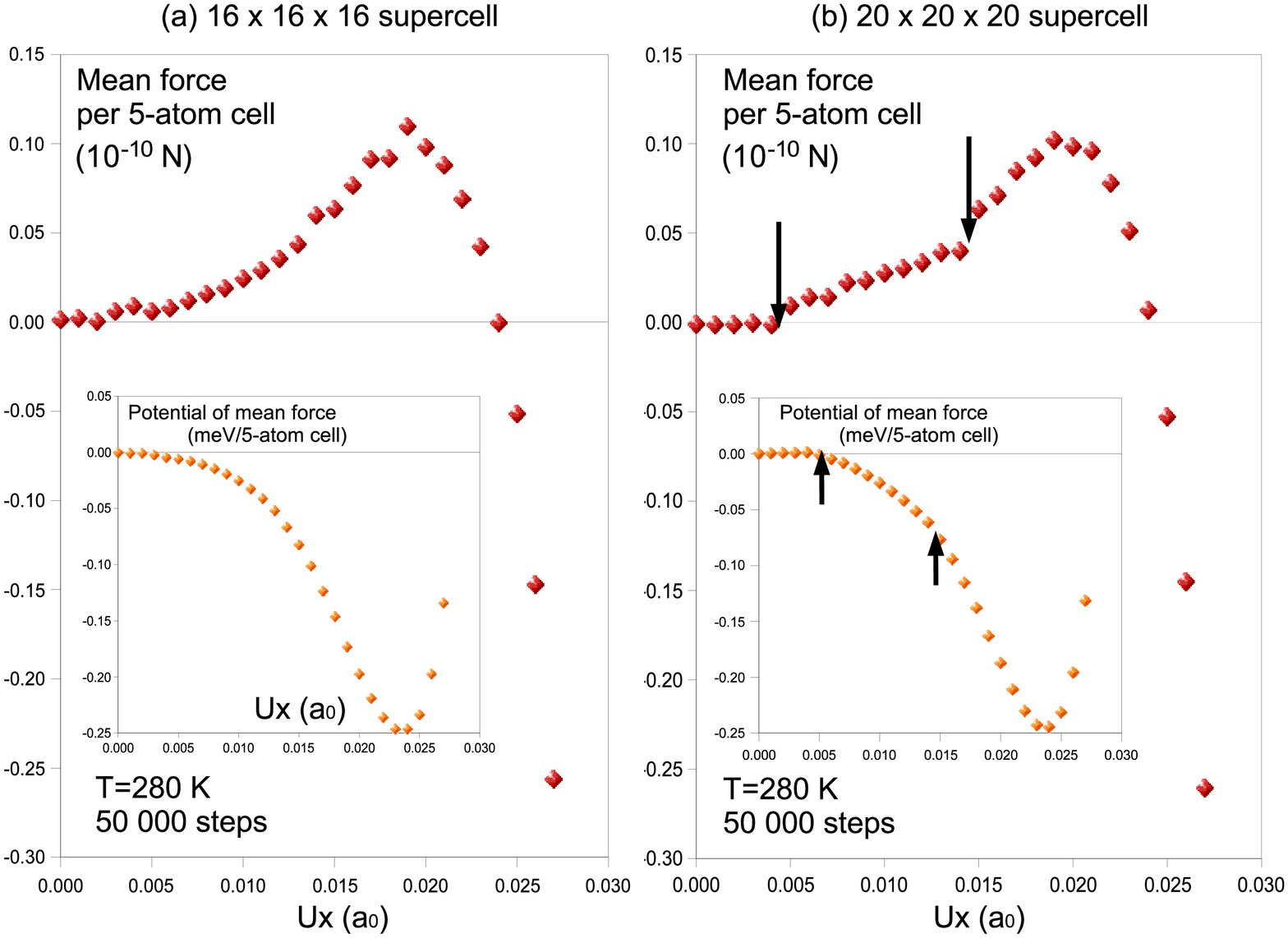}}}
    \par}
     \caption{{\small Mean force as a function of mean local mode (arb. units) and potential of mean force (obtained by integration of minus the mean force). (a): 16 $\times$ 16 $\times$ 16 supercell, average over 50 000 steps (b): 20 $\times$ 20 $\times$ 20 supercell, average over 50 000 steps.}}
    \label{supercell_effect}
\end{figure}

The effect of the supercell size is tested by performing simulations at T=280 K in a 16 $\times$ 16 $\times$ 16 and in a 20 $\times$ 20 $\times$ 20 supercell. The total mean force obtained as a function of mean local mode, as well as the corresponding potential of mean force (free energy) are shown on Fig.~\ref{supercell_effect}.

In the case of the 16 $\times$ 16 $\times$ 16 supercell, the agreement is very good with respect to the 12 $\times$ 12 $\times$ 12 cell: the mean force and free energy curves are very close to those presented above. Plotting the local mode distribution as well as the strain tensor component evolution indicates the absence of phase separation at low order parameter and confirms that interfacial contributions to the free energy are negligible, if any.

\begin{figure}[htbp]
    {\par\centering
    {\scalebox{0.65}{\includegraphics{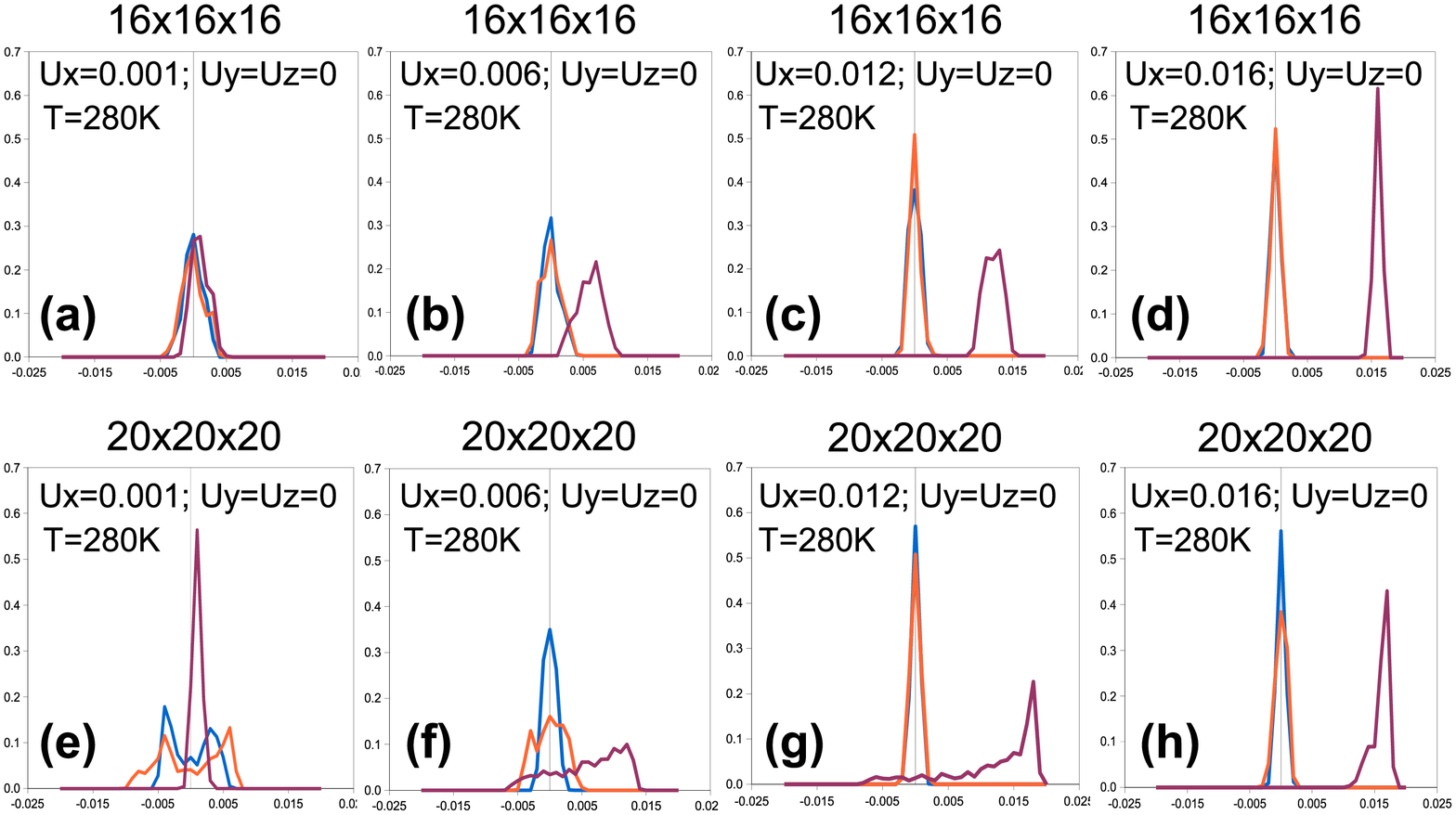}}}
    \par}
     \caption{{\small Distribution of the time-averaged components of the local modes (arb. units), for 4 constrained values of the order parameter, at T=280 K. Panels a,b,c,d: 16 $\times$ 16 $\times$ 16 supercell. Panels e,f,g,h: 20 $\times$ 20 $\times$ 20 supercell. The purple (resp. blue and orange) lines refer to the $x$ (resp. $y$ and $z$) component. The local modes are in a$_0$ (lattice constants) units.}}
    \label{supercell_effect2}
\end{figure}

However, in the 20 $\times$ 20 $\times$ 20 supercell, phase separation starts to manifest at low order parameter, yielding a characteristic trend on the mean force and free energy curve: some "plateaus" do appear in the mean force curve. This can be observed on the local mode distribution (Fig.~\ref{supercell_effect2}) as well as on the strain tensor components. Consequently, the mean force and the free energy deviate slightly from the two curves obtained with smaller supercells. More precisely, three regions can be distinguished in that case, separated on Fig.~\ref{supercell_effect} by black arrows: at low order parameter (close to zero), the mean force and the free energy are flat, indicating pseudo-domains with roughly opposite polarizations in the directions perpendicular to $x$ ($x$ is the direction of the fixed order parameter). Then at $u_x \approx$ 0.005 a$_0$, the mean force increases but through a linear curve (instead of a convex curve as in the smaller supercells). From Fig.~\ref{supercell_effect2}, this situation corresponds to pseudo-domains with polarization oriented along $x$, very inhomogeneous (Fig.~\ref{supercell_effect2}, panels (f) and (g)), but the polarization in the perpendicular directions is maintained to its fixed value. Then at $u_x \approx$ 0.015 a$_0$, the distributions of the various components of the time-averaged local modes recover single-peak features, and the mean force recovers similar values as in smaller supercells.

\begin{figure}[htbp]
    {\par\centering
    {\scalebox{0.65}{\includegraphics{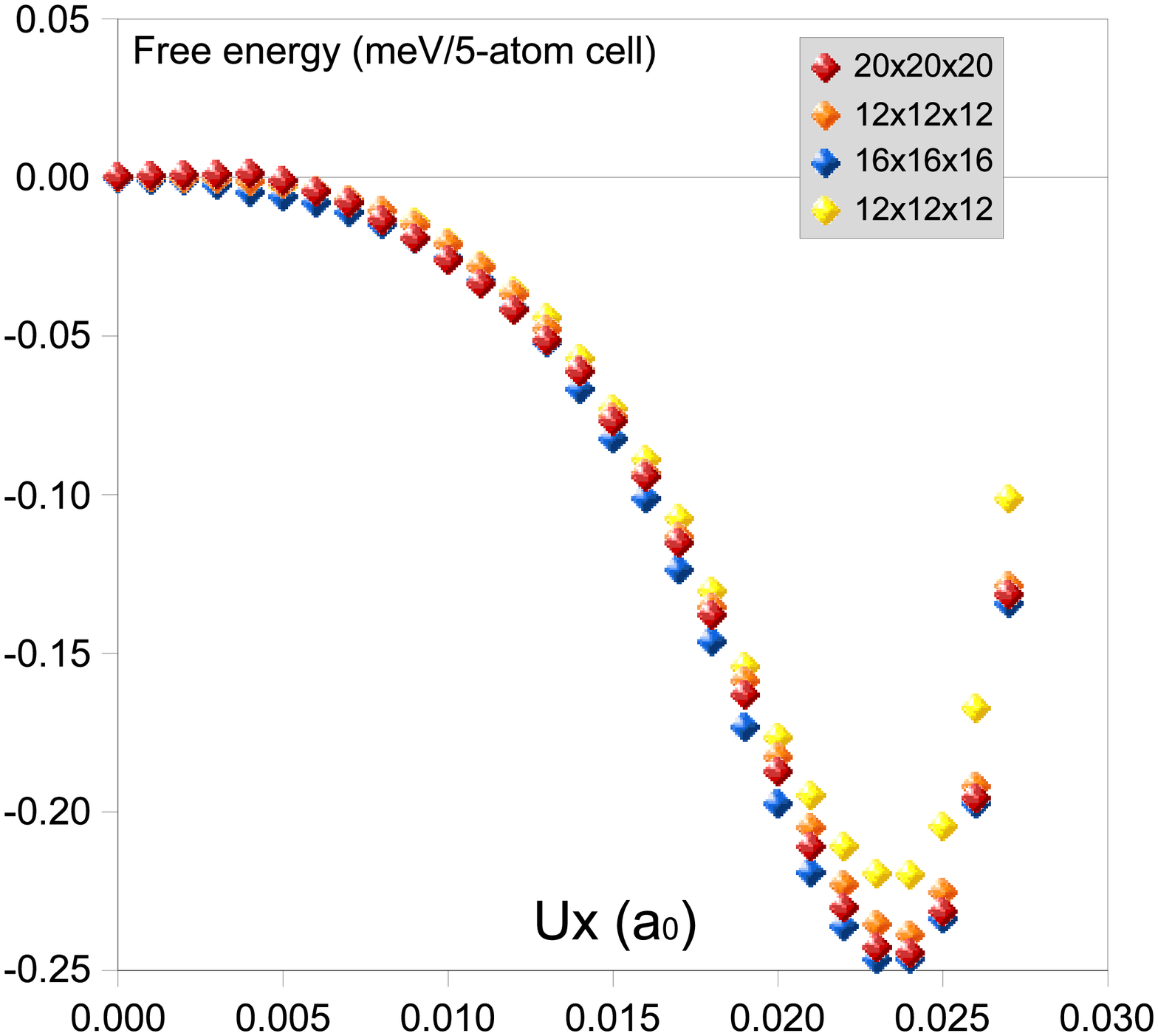}}}
    \par}
     \caption{{\small Free energies as a function of mean local mode (meV/5-atom cell), as calculated for the three supercell sizes. Yellow diamonds: 12 $\times$ 12 $\times$ 12 (50 000 steps);
     Orange diamonds: 12 $\times$ 12 $\times$ 12 (500 000 steps);
     Red diamonds: 20 $\times$ 20 $\times$ 20 (50 000 steps);
     Blue diamonds: 16 $\times$ 16 $\times$ 16 (50 000 steps). }}
    \label{supercell_effect3}
\end{figure}

The free energy, as a consequence, even though the agreement with previous curves is still good (Fig.~\ref{supercell_effect3}), tends to differ slightly. This trend is probably enstrengthed in larger supercells.

Too large supercells thus lead to low order parameter states in which inhomogeneous configurations are dominating the sampling, and the free energies obtained are consequently not smooth in these regions. In the following, we use the 12 $\times$ 12 $\times$ 12 supercell to compute the free energy around T$_c$.

From the previous tests, we may assume that at T=280 K, the supercell sizes L=12 and L=16 are lower than some characteristic length of the system that can be compared to or interpreted as a correlation length. L=20 lattice constants is probably close to this correlation length. We show in the following that the computed free energies for L=12 are fully comparable to the phenomenological Landau free energies of the litterature.
Supercell sizes ranging from L=12 to L=20 provide free energies (as a function of order parameter) that are quasi-similar at this temperature. Note that when approaching T$_c$ from below, the correlation length increases and the free energy curves computed around T$_c$ with L=12 are thus probably more sensitive to finite-size effects.

\subsection{States of constrained polarization at lower temperature}
\label{temperature}
\begin{figure}[htbp]
    {\par\centering
    {\scalebox{0.65}{\includegraphics{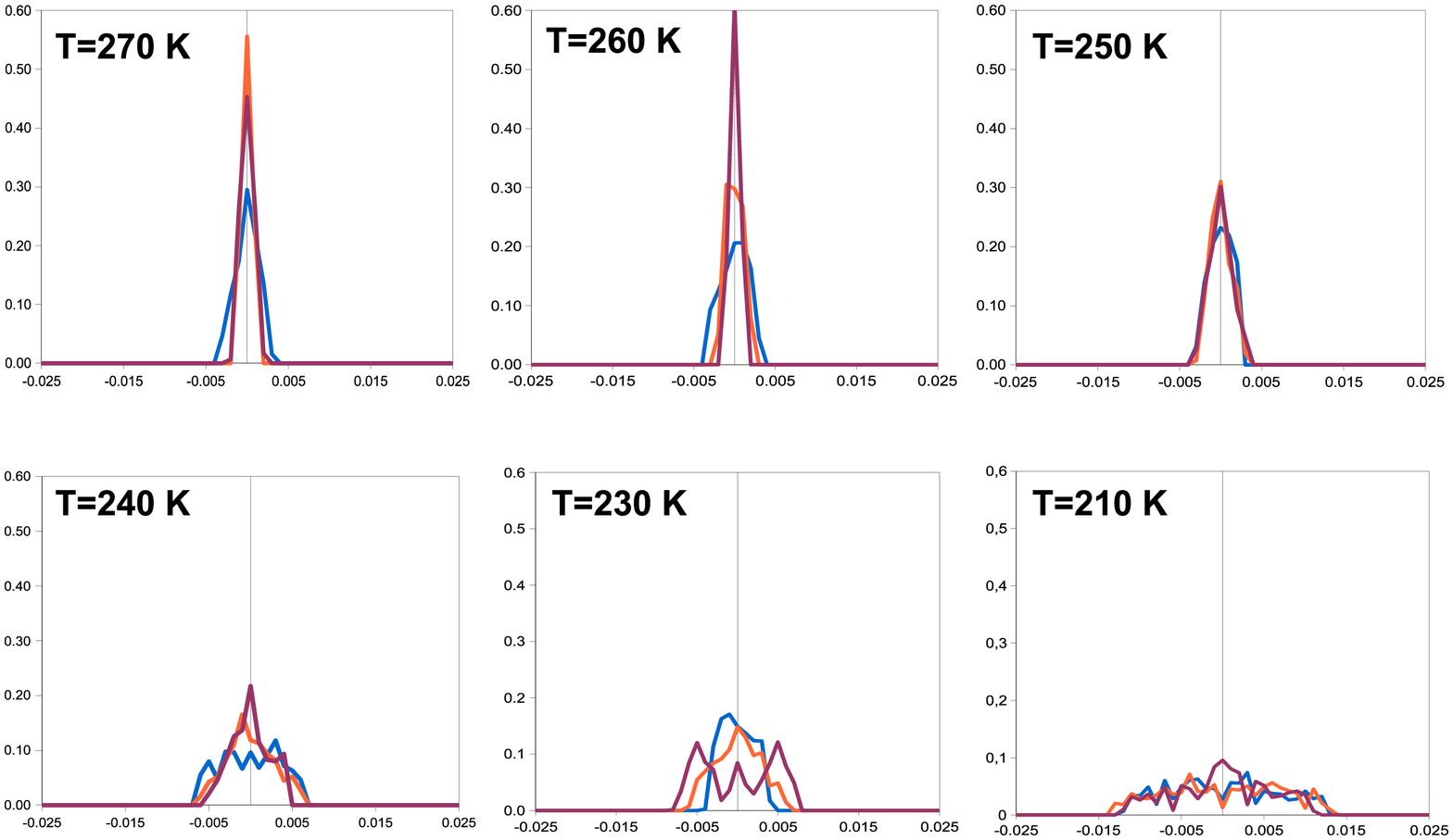}}}
    \par}
     \caption{{\small Distribution of the time-averaged local modes in a 12 $\times$ 12 $\times$ 12 supercell after 100 000 steps for a constrained value of the order parameter ($u_x=0,~u_y=0,~u_z=0$). The unit is the same as in the previous figures. The local modes are in a$_0$ (lattice constants) units.}}
    \label{low_T}
\end{figure}

When the temperature decreases below 280 K, the states of constrained polarization at low order parameter become progressively more and more inhomogeneous, especially for $T \leq$ 240 K. This is illustrated on Fig.~\ref{low_T}, that represents the distribution of the time-averaged local modes in a L=12 supercell between T=270 K and  T=210 K.

\begin{figure}[htbp]
    {\par\centering
    {\scalebox{0.35}{\includegraphics{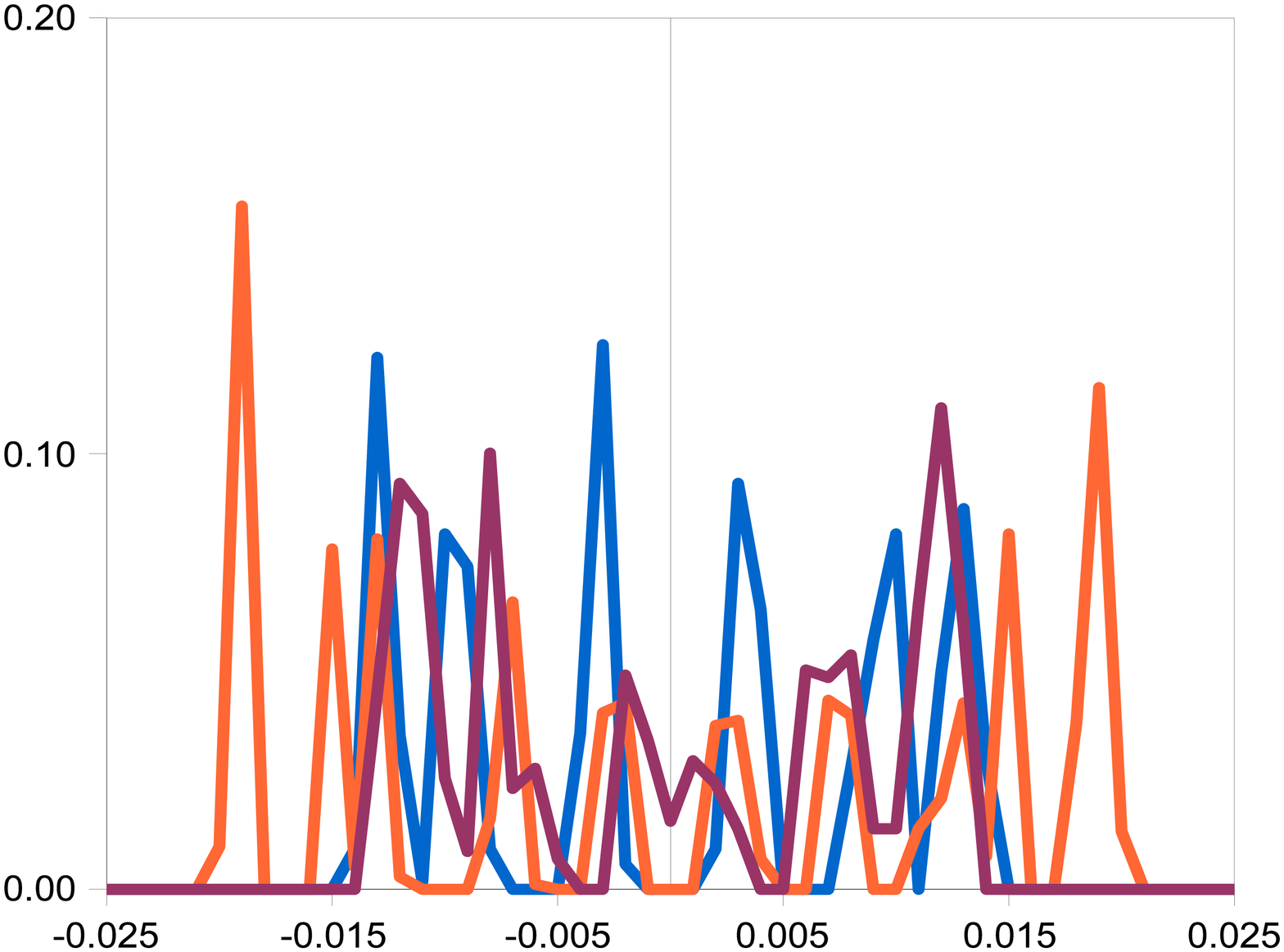}}}
    \par}
     \caption{{\small Distribution of the time-averaged local modes in a 12 $\times$ 12 $\times$ 12 supercell after 350 000 steps at T=200 K for a constrained value of the order parameter ($u_x=0,~u_y=0,~u_z=0$). The unit is the same as in the previous figures. The local modes are in a$_0$ (lattice constants) units.}}
    \label{low_T2}
\end{figure}

At very low temperature and low order parameter, this distribution exhibits several peaks that reflect the appearance of very inhomogeneous states (Fig.~\ref{low_T2}).

\section{Results}

\subsection{Computation of the free energy around T$_c$ using a 12 $\times$ 12 $\times$ 12 supercell}

\begin{figure}[htbp]
    {\par\centering
    {\scalebox{0.38}{\includegraphics{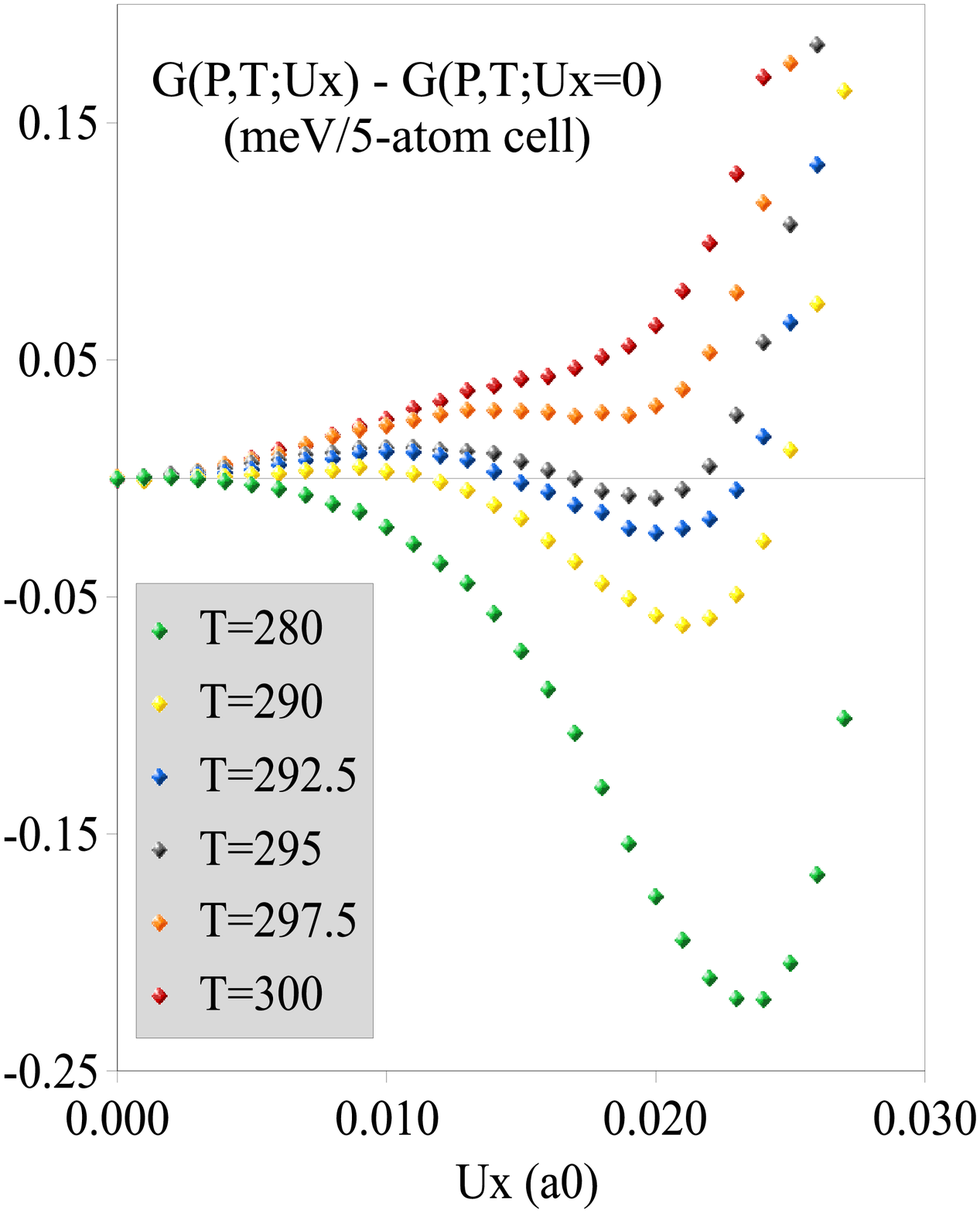}}}
    \par}
     \caption{{\small Gibbs free energy as a function of $u_x$ (with $u_y=u_z=0$, {\it i.e.} along [100]) for five temperatures around the paraelectric-ferroelectric phase transition. $u_x$ is in units a$_0$ (= 7.46 bohr). The supercell is 12 $\times$ 12 $\times$ 12.}}
    \label{figure1}
\end{figure}

The Gibbs free energy curves as a function of $u_x$, computed in a 12 $\times$ 12 $\times$ 12 supercell for T=280, 290, 292.5, 295, 297.5 and 300 K are shown on Fig.~\ref{figure1}. They correspond very well to the profiles expected from Landau theory for a first-order phase transition, with a free energy barrier separating the paraelectric phase from the ferroelectric phase for a few K below and above T$_c$. From these curves, we can localize the Curie temperature between 295 and 297.5 K (I$\tilde{n}$iguez {\it et al} localize it at 297 K with the same hamiltonian\cite{iniguez2001}). The free energy barrier separating the cubic paraelectric phase and the tetragonal ferroelectric phase at T$_c$ is $\Delta G \approx$ 0.012-0.015 meV/5-atom cell. The order of magnitude of this small barrier is in good agreement with that given by phenomenological Landau Gibbs free energies such as that of Wang {\it et al}\cite{wang2007} for which $\Delta G \approx$ 0.01 meV/5-atom cell at T$_c$. 

\subsection{Comparison with phenomenological potentials}
We now compare the results with classical phenomenological Landau potentials, that assume in particular a linear dependence of the quadratic coefficient with temperature. In the litterature, these potentials can be found as Gibbs free energies for the stress-free crystal, under the form of polynomial functions up to 6$^{th}$ or 8$^{th}$ order in the $P_x$, $P_y$ and $P_z$. For example, the 8$^{th}$ order development writes:

\begin{eqnarray*}
&& G(T;\vec P) = \alpha_{1}(P_x^2 + P_y^2 + P_z^2) + \alpha_{11} (P_x^4 + P_y^4 + P_z^4)  \\
&& + \alpha_{12}(P_x^2 P_y^2 + P_x^2 P_z^2 + P_y^2 P_z^2) + \alpha_{111} (P_x^6 + P_y^6 + P_z^6) \\
&& + \alpha_{112}(P_x^2(P_y^4+P_z^4) + P_y^2(P_x^4+P_z^4) + P_z^2(P_x^4+P_y^4)) \\
&& +\alpha_{123}P_x^2 P_y^2 P_z^2 + \alpha_{1111}(P_x^8 + P_y^8 + P_z^8)  \\
&& +\alpha_{1112}(P_x^6(P_y^2+P_z^2) + P_y^6(P_x^2+P_z^2) + P_z^6(P_x^2+P_y^2))  \\
&& + \alpha_{1122}(P_x^4 P_y^4 + P_x^4 P_z^4 + P_y^4 P_z^4)  \\
&& +\alpha_{1123}(P_x^4 P_y^2 P_z^2 + P_y^4 P_x^2 P_z^2 + P_z^4 P_x^2 P_y^2)
\end{eqnarray*}

The free energy along the $x$ axis ($P_x \neq 0$, $P_y = P_z = 0$) writes:
\begin{equation}
 G(T;\vec P) = \alpha_{1}P_x^2  + \alpha_{11} P_x^4  + \alpha_{111} P_x^6 +  \alpha_{1111}P_x^8 
\end{equation}

A fit of our free energy curves by 6$^{th}$-order or 8$^{th}$-order polynomial functions (without odd-order term) $aP^2 + bP^4 + cP^6 (+dP^8)$ should directly provide coefficients comparable to the $\alpha_{1}$, $\alpha_{11}$, $\alpha_{111}$ and $\alpha_{1111}$ of the litterature. Thus we fit these curves by such polynomial functions and extract the coefficients (the volume variation along the curve is accounted for when performing the calculation).

\begin{figure}[htbp]
    {\par\centering
    {\scalebox{0.30}{\includegraphics{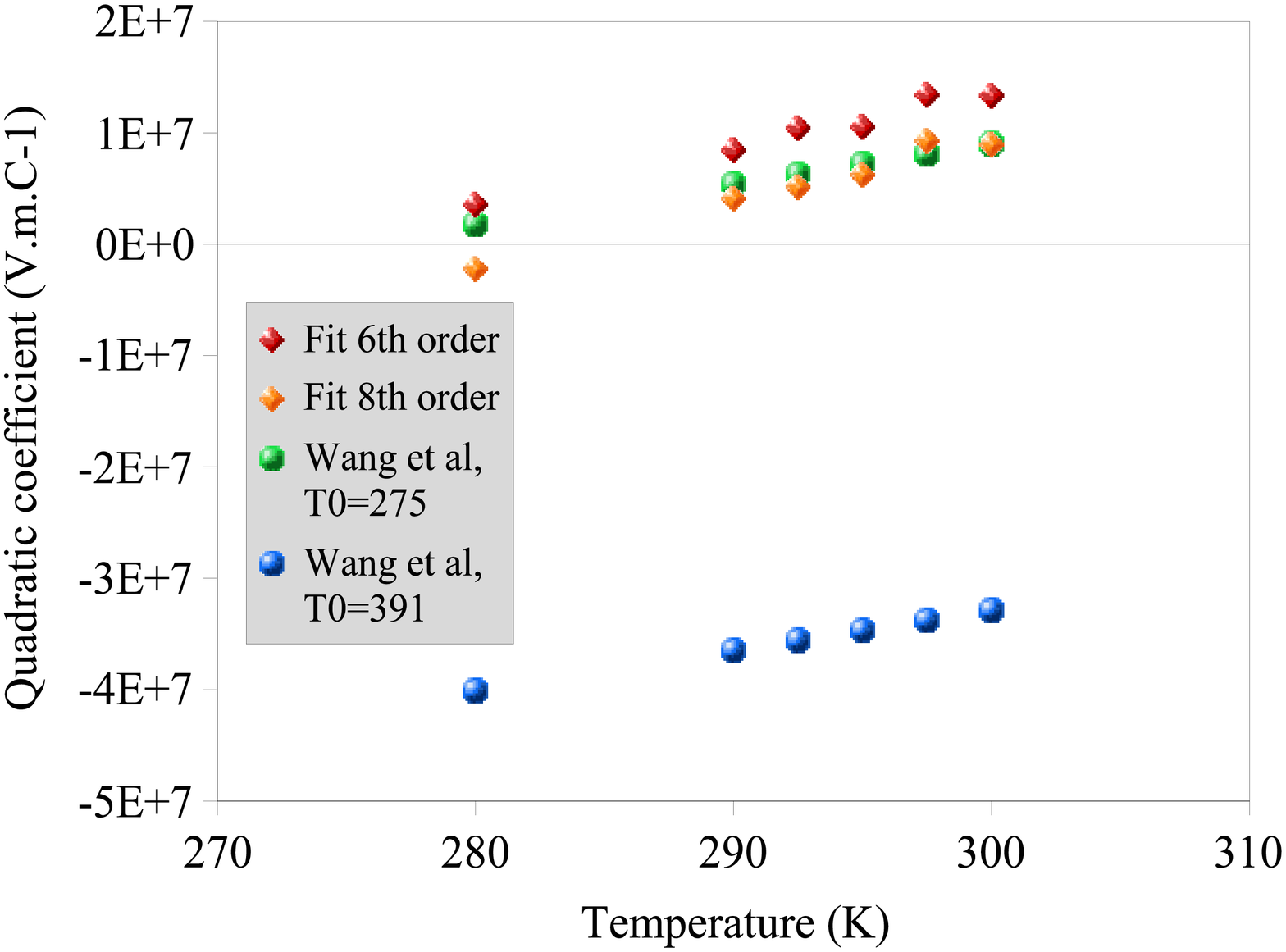}}}
    \par}
     \caption{{\small Temperature evolution of the quadratic coefficient (between 280 and 300 K), according to a fit of the computed free energy on a 6$^{th}$ order (red diamonds) or 8$^{th}$ order (orange diamonds) polynomial. The circles refer to the Landau phenomenological potential of Ref.~\onlinecite{wang2007}.}}
    \label{figure3}
\end{figure}

First the computed free energy curves fit very well by such functions. The quadratic term obtained by fitting is plotted on Fig.~\ref{figure3} as a function of T for the two fits (6$^{th}$ and 8$^{th}$ order). On the same figure, we plot the quadratic term of the phenomenological potential of Wang {\it et al}\cite{wang2007} $\alpha_1$ = 3.61$\times$10$^5$(T-T$_0$) (V.m.C$^{-1}$). In Landau theory for first order phase transitions, T$_0$ is, just below T$_c$, the temperature above which the paraelectric state ($\vec P = \vec 0$) becomes a (local) minimum of $G$. Wang {\it et al} set this value at 391 K, just below the real Curie temperature. In our case, T$_0$ has to be readjusted below the Curie temperature of the Effective Hamiltonian. The best agreement with the 8$^{th}$ order polynomial fit is obtained for T$_0$=275 K (green circles on Fig.~\ref{figure3}), which is indeed just below the Curie temperature of the hamiltonian ($\approx$ 297 K). For this value of T$_0$, the agreement between the phenomenological potential of Wang {\it et al}\cite{wang2007} and our work is very good. The temperature dependence of the quadratic term fitted is linear on the small range of temperature considered here (280-300 K). This would probably not be the case if the study was extended to lower temperatures\cite{iniguez2001}.

\begin{figure}[htbp]
    {\par\centering
    {\scalebox{0.40}{\includegraphics{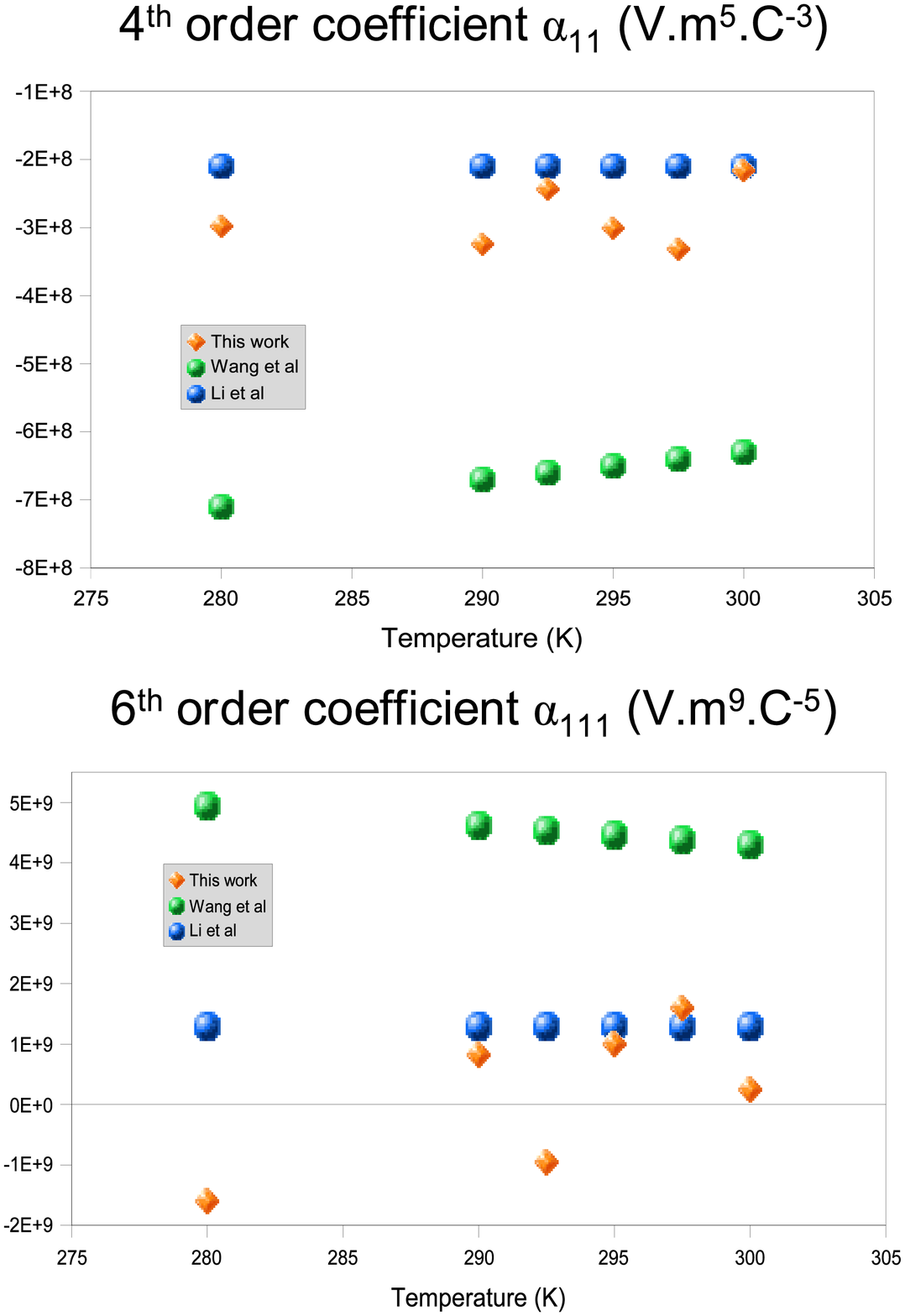}}}
    \par}
     \caption{{\small Temperature evolution of the quartic ($\alpha_{11}$) and sixth order ($\alpha_{111}$) coefficients, as obtained from this work (orange diamonds) and as provided by Wang {\it et al} (green circles) and Li {\it et al} (blue circles).}}
    \label{figure4}
\end{figure}

\begin{figure}[htbp]
    {\par\centering
    {\scalebox{0.40}{\includegraphics{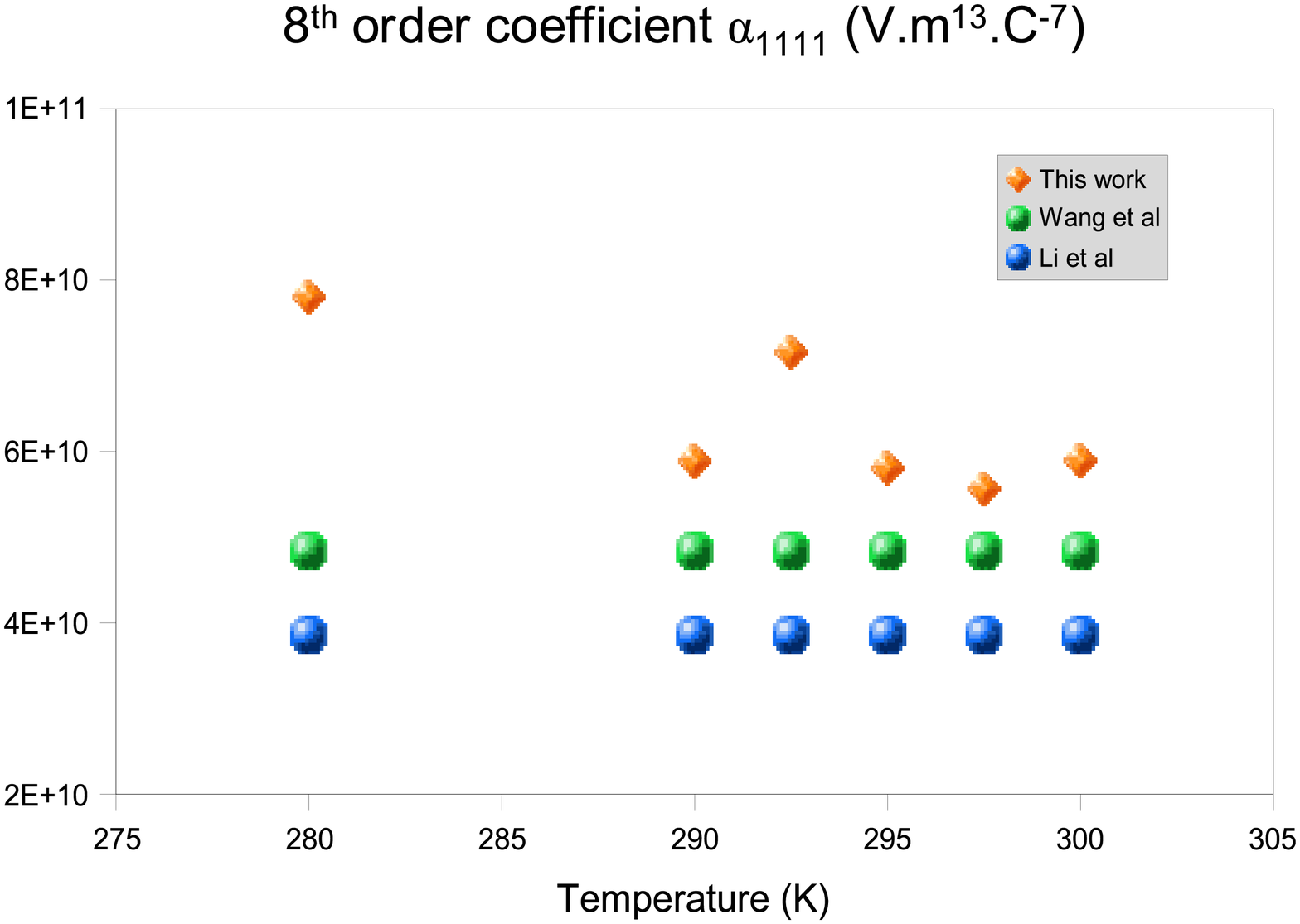}}}
    \par}
     \caption{{\small Temperature evolution of the eight order coefficient $\alpha_{1111}$, as obtained from this work (orange diamonds) and as provided by Wang {\it et al} (green circles) and Li {\it et al} (blue circles).}}
    \label{figure5}
\end{figure}

The temperature evolutions of the other coefficients are shown on Figs.~\ref{figure4} (quartic and sixth order coefficient) and ~\ref{figure5} (eight-order coefficient) in the case of a fit by an 8$^{th}$ order polynomial function. We find an excellent agreement close to T$_c$ with the two phenomenological potentials of Wang {\it et al}\cite{wang2007} and Li {\it et al}\cite{li2005}, for the three coefficients $\alpha_{11}$, $\alpha_{111}$ and $\alpha_{1111}$. 

\subsection{Natural decomposition of the free energy}

Since the effective hamiltonian decomposes into various contributions\cite{zhong95} (in particular onsite, short-range, dipole-dipole, elastic-local mode interaction), the force acting on the $i^{th}$ local mode $\vec f^{lm}_i$ also naturally decomposes into:

\begin{equation}
\nonumber
\vec f^{lm}_i = - \frac{\partial H^{eff}}{\partial \vec u_i}
= \vec f^{self}_i + \vec f^{short}_i + \vec f^{dpl}_i + \vec f^{int}_i,
\end{equation}

with obvious notations. This leads to a natural decomposition of the free energy $\Delta \tilde{F}(N,\left\{ \eta \right\},T;\vec u) = \tilde{F}(N,\left\{ \eta \right\},T;\vec u) - \tilde{F}_0(N,\left\{ \eta \right\},T)$ as:

\begin{equation}
\Delta \tilde{F}(N,\left\{ \eta \right\},T;\vec u) = \Delta \tilde{F}^{self}(N,\left\{ \eta \right\},T;\vec u)+ \Delta \tilde{F}^{short}(N,\left\{ \eta \right\},T;\vec u)+ \Delta \tilde{F}^{dpl}(N,\left\{ \eta \right\},T;\vec u)+ \Delta \tilde{F}^{int}(N,\left\{ \eta \right\},T;\vec u),
\end{equation}

which extends easily to the Gibbs free energy:

\begin{equation}
\Delta \tilde{G}(N,\left\{ \sigma \right\},T;\vec P) = \Delta \tilde{G}^{self}(N,\left\{ \sigma \right\},T;\vec P)+ \Delta \tilde{G}^{short}(N,\left\{ \sigma \right\},T;\vec P)+ \Delta \tilde{G}^{dpl}(N,\left\{ \sigma \right\},T;\vec P)+ \Delta \tilde{G}^{int}(N,\left\{ \sigma \right\},T;\vec P),
\end{equation}

\begin{figure}[htbp]
    {\par\centering
    {\scalebox{0.30}{\includegraphics{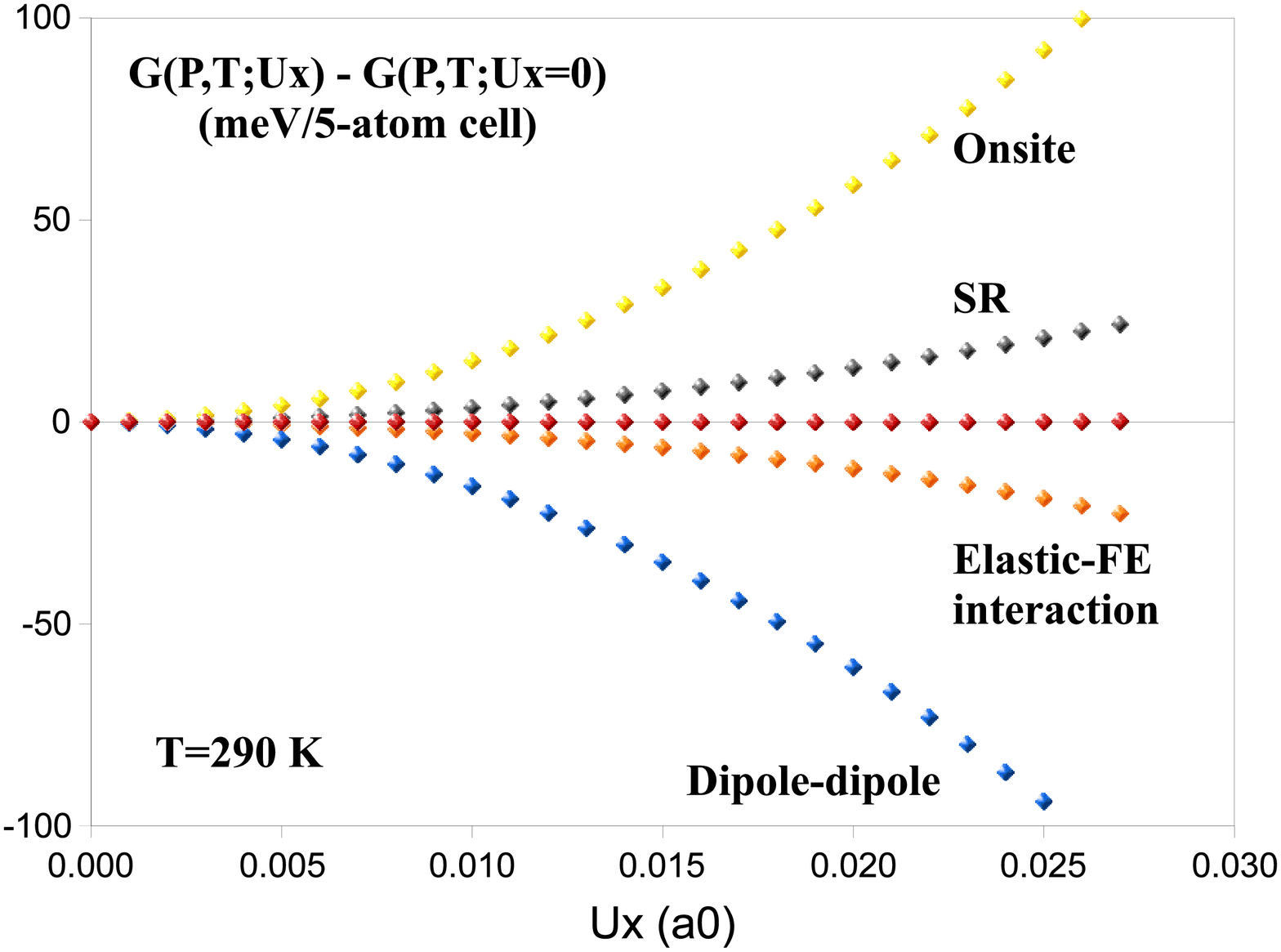}}}
    \par}
     \caption{{\small Decomposition of the free energy (red) into its 4 contributions at T=290 K, as a function of $U_x$ (with $U_y=U_z=0$) ({\it i.e.} along [100]). Each of the 4 contributions is at least 2 orders of magnitude higher than the free energy itself, that appears therefore completely flat (red diamonds).}}
    \label{figure2}
\end{figure}

First, we notice that in the vicinity of the phase transition, these contributions are much higher (at least two orders of magnitude) than the free energy itself, confirming the well-known trend, according to which the existence of ferroelectricity is a very delicate balance between large destabilizing contributions (here short-range and onsite) and large stabilizing ones (dipole-dipole and elastic-local mode interaction), as illustrated on Fig.~\ref{figure2}.

\subsubsection{Short-range and electrostatic parts}
Examining the various contributions as a function of $\vec u$ (mean local mode), we notice something {\it a priori} surprising: among the four contributions, two of them (the short-range and the dipole-dipole) are  independent on the temperature. This would not be the case if these contributions were expressed as a function of $\vec P$ instead of $\vec u$ (because the volume depends on $\vec u$, see Fig.~\ref{figure0}). In fact, this is simply due to the fact that the corresponding parts of the Effective Hamiltonian are quadratic in the local modes: 

\begin{equation}
\nonumber
E^{short} = \frac{1}{2} \sum_{i,j,\alpha,\beta} J_{ij}^{\alpha,\beta} u_{i,\alpha} u_{j,\beta} 
 \Rightarrow f^{short}_{i,\alpha} = -\frac{\partial E^{short}}{\partial u_{i,\alpha}}  =
- \sum_{j,\beta} J_{ij}^{\alpha,\beta} u_{j,\beta},
\end{equation}

Thus the forces vary linearly with the local modes, and their thermal average (in any ensemble) is

\begin{equation}
\nonumber
\left\langle f^{short}_{i,\alpha}\right\rangle = - \sum_{j,\beta} J_{ij}^{\alpha,\beta} \left\langle u_{j,\beta} \right\rangle
\end{equation}

Under fixed $\vec u$, and under the hypothesis that the system is homogeneous ($\forall j,\beta,~ \left\langle u_{j,\beta} \right\rangle =  u_{\beta}$):

\begin{equation}
\nonumber
\left\langle f^{short}_{i,\alpha}\right\rangle = - \sum_{j,\beta} J_{ij}^{\alpha,\beta} u_{\beta}
\end{equation}

Thus 

\begin{equation}
\nonumber
\sum_i \left\langle f^{short}_{i,\alpha}\right\rangle = - \sum_{i,j,\beta} J_{ij}^{\alpha,\beta} u_{\beta}
\end{equation}

\begin{equation}
\nonumber
\frac{\partial \tilde{G}^{short}}{\partial u_{\alpha}} = - \sum_i \left\langle f^{short}_{i,\alpha}\right\rangle_{NPT}
= \sum_{\beta} \left\{ \sum_{i,j} J_{ij}^{\alpha,\beta}\right\} u_{\beta}
\end{equation}

Let us introduce the 3$\times$3 matrix: $J_{\alpha,\beta} = \sum_{i,j} J_{ij}^{\alpha,\beta}$.

\begin{equation}
\frac{\partial \tilde{G}^{short}}{\partial u_{\alpha}} = \sum_{\beta}  J_{\alpha,\beta}  u_{\beta}
\end{equation}

The same arguments stand of course in the canonical ensemble:
\begin{equation}
\frac{\partial \tilde{F}^{short}}{\partial u_{\alpha}} = \sum_{\beta}  J_{\alpha,\beta}  u_{\beta}
\end{equation}

It follows:

\begin{equation}
\Rightarrow \Delta \tilde{G}^{short} = \Delta \tilde{F}^{short} = \frac{1}{2} \sum_{\alpha,\beta}  J_{\alpha,\beta} u_{\alpha} u_{\beta}
\end{equation}

The SR contribution to the free energy is thus independent on the temperature and quadratic in the mean local mode.

For the dipole-dipole part, we have also:
\begin{equation}
\frac{\partial \tilde{G}^{dpl}}{\partial u_{\alpha}} = \frac{\partial \tilde{F}^{dpl}}{\partial u_{\alpha}}
= 2 \sum_{\beta}  Q_{\alpha,\beta}  u_{\beta}
\end{equation}

It follows:

\begin{equation}
\Rightarrow \Delta \tilde{G}^{dpl} = \Delta \tilde{F}^{dpl} = \sum_{\alpha,\beta}  Q_{\alpha,\beta} u_{\alpha} u_{\beta}
\end{equation}

with $Q_{\alpha,\beta} = \sum_{i,j} Q_{ij}^{\alpha,\beta}$, the Q matrix defined in Ref.~\onlinecite{zhong95} to compute the electrostatic energy.

The previous equations allow to compute directly the short-range and dipole-dipole part of the free energy from the coefficients of the effective hamiltonian without simulation. These parts are quadratic in $\vec u$ and independent on the temperature.

However, we stress that these two parts are independent on the temperature when expressed as a function of the mean local mode $\vec u$. If they are expressed as a function of the polarization $\vec P$, a temperature dependence appears, related to the variation of the cell volume ($\Omega/N$) with temperature: $\vec P = N Z^{*}/ \Omega(P,T,\vec u) \vec u$.

\subsubsection{Onsite part and elastic-local mode interaction part}
The temperature dependence of the free energy, when expressed as a function of $\vec u$, thus originates only in the onsite and elastic-local mode interaction parts of the Effective Hamiltonian. Let us examine the onsite energy: the corresponding part of the Effective Hamiltonian is local and has an harmonic part and an anharmonic part:
\begin{equation}
\nonumber
E^{self} = \sum_{i,\alpha,\beta} A_{\alpha,\beta} u_{i,\alpha} u_{i,\beta} + E^{self}_{anharm}
\end{equation}

From the arguments given above, only the anharmonic part is likely to generate a temperature dependence of the corresponding part of the free energy:
\begin{equation}
\nonumber
\frac{\partial \tilde{G}^{self}}{\partial u_{\alpha}} = N \sum_{\beta}  A_{\alpha,\beta}  u_{\beta} + 
\frac{\partial \tilde{G}^{self}_{anharm}}{\partial u_{\alpha}}
\end{equation}

We have also:

\begin{equation}
\nonumber
\frac{\partial \tilde{F}^{self}}{\partial u_{\alpha}} = N \sum_{\beta}  A_{\alpha,\beta}  u_{\beta} + 
\frac{\partial \tilde{F}^{self}_{anharm}}{\partial u_{\alpha}} 
\end{equation}

Thus,
\begin{equation}
\nonumber
\Rightarrow \Delta \tilde{G}^{self} =  N \sum_{\alpha,\beta}  A_{\alpha,\beta} u_{\alpha} u_{\beta} + \Delta \tilde{G}^{self}_{anharm}
\end{equation}
and
\begin{equation}
\nonumber
\Rightarrow \Delta \tilde{F}^{self} =  N \sum_{\alpha,\beta}  A_{\alpha,\beta} u_{\alpha} u_{\beta} + \Delta \tilde{F}^{self}_{anharm}
\end{equation}

With the notations of Ref.~\onlinecite{zhong95}, we have:
\begin{equation}
\Delta \tilde{F}^{self} = N\kappa_2 \left\| \vec u \right\|^2 + \Delta \tilde{F}^{self}_{anharm}
\end{equation}
\begin{equation}
\Delta \tilde{G}^{self} = N\kappa_2 \left\| \vec u \right\|^2 + \Delta \tilde{G}^{self}_{anharm}
\end{equation}

The contributions from anharmonic terms might contain also {\it a priori} an harmonic contribution to the free energy. Although this term in the Effective Hamiltonian only has quartic contributions, we have shown above that a fitting of the free energy on polynomial functions may contain higher-order terms.

Finally, we focus on the elastic-local mode interaction energy:
\begin{equation}
\nonumber
E^{int} = \frac{1}{2} \sum_{l,i,\alpha,\beta} B_{l,\alpha,\beta} \eta_l(i) u_{i,\alpha} u_{i,\beta},
\end{equation}

with $\eta_l(i) = \eta_l^H + \eta_l^I(i)$ (homogeneous strain + inhomogeneous strain).
The same treatment can not be applied here due to presence of the inhomogeneous strain. This contribution is thus {\it a priori} dependent on the temperature and this is observed in our simulations.

\section{Discussion and conclusion}
In this work, we have computed the free energy of barium titanate as a function of polarization directly from an effective hamiltonian. We have used Molecular Dynamics combined with Thermodynamic Integration. This is equivalent to define the free energy as the potential of the mean force acting on the polarization. One obtains, for a given temperature, the difference of free energy between $\vec P = \vec 0$ and $\vec P$. The simulations can be performed under fixed volume/strain or fixed pressure/stress conditions, which gives access to Helmholtz free energies or to Gibbs free energies. The gradient of the free energy is related to the thermal average of the total forces, computed under the constraint of fixed polarization. From the decomposition of the forces into the different contributions coming from the Effective Hamiltonian, a natural decomposition of the free energy of a ferroelectric into four parts (onsite, dipole-dipole, short-range and elastic-local mode interaction) has been suggested. In particular we have shown that the dipole-dipole and short-range contributions to the free energy are, in the framework of the Effective Hamiltonian of Zhong {\it et al}\cite{zhong95}, independent on the temperature when expressed as a function of $\vec u$ instead of $\vec P$ and are quadratic in $\vec u$.

The computation has been performed for a set of temperatures around T$_c$ and by using a 12 $\times$ 12 $\times$ 12 supercell. Although the comparison of such a free energy with a Landau potential is still controversial\cite{troster2005}, a very good agreement has been found with Landau phenomenological potentials of the litterature\cite{wang2007,li2005} aroud T$_c$. We have checked that in the range of temperature considered, phase separation at low order parameter does not occur within this supercell. Moreover we have shown, by systematically increasing the supercell size and the number of time steps of the simulation, that the computed free energy does not contain significant interfacial contributions and can be considered as a volume (extensive) quantity. We think that these techniques can be useful to approach the Landau free energy of a ferroelectric near the Curie temperature and by using atomic-scale simulations.

As reminded in the introduction, the concept of Landau free energy requires the choice of a spatial averaging length $L$ to define the order parameter as a smooth and continuous physical quantity. If this averaging length is chosen above the correlation length $\xi$, some local modes might be unsufficiently correlated within the averaging region and microscopic configurations with phase separation are likely to appear frequently in the sampling when a low value of the (averaged) order parameter is fixed. These phenomena, already reported in the framework of the $\phi^4$ model, have been also observed in the present case at low temperature or when the supercell size is very large.
Such considerations lead to the concept of coarse-grained free energies\cite{troster2005} and elegant approaches have been proposed to circumvent the corresponding difficulties\cite{troster2007}.

However, as reminded also, the definition of the correlation volume is rather complex in ferroelectric systems due to the very anisotropic nature of the correlations. Further investigation of the notion of correlation length/volume in ferroelectrics should be of primary interest to enlight the concept of Landau free energy in those materials.

\end{document}